\documentclass[aps,pre,onecolumn,superscriptaddress]{revtex4-2}

\usepackage{graphicx}
\usepackage{color}

\begin{document}

\title{Lethal DNA damages caused by ion-induced shock waves in cells}

\author{Ida Friis}
\affiliation{Department of Physics, Chemistry and Pharmacy, University of Southern Denmark, Campusvej 55, 5230 Odense M, Denmark}

\author{Alexey V. Verkhovtsev}
\altaffiliation{On leave from Ioffe Institute, Polytekhnicheskaya 26, 194021 St. Petersburg, Russia}
\affiliation{MBN Research Center, Altenh\"oferallee 3, 60438 Frankfurt am Main, Germany}

\author{Ilia A. Solov'yov}
\email[]{ilia.solovyov@uni-oldenburg.de}
\altaffiliation{On leave from Ioffe Institute, Polytekhnicheskaya 26, 194021 St. Petersburg, Russia}
\affiliation{Department of Physics, Carl von Ossietzky Universit\"at Oldenburg, Carl-von-Ossietzky-Str. 9-11, 26129 Oldenburg, Germany}

\author{Andrey V. Solov'yov}
\altaffiliation{On leave from Ioffe Institute, Polytekhnicheskaya 26, 194021 St. Petersburg, Russia}
\affiliation{MBN Research Center, Altenh\"oferallee 3, 60438 Frankfurt am Main, Germany}


\begin{abstract}
The elucidation of fundamental mechanisms underlying ion-induced radiation damage of biological systems is crucial for advancing radiotherapy with ion beams and for radiation protection in space. The study of ion-induced biodamage using the phenomenon-based MultiScale Approach to the physics of radiation damage with ions (MSA) has led to the prediction of nanoscale shock waves created by ions in a biological medium at the high linear energy transfer (LET). The high-LET regime corresponds to the keV and higher energy losses by ions per nanometer, which is typical for ions heavier than carbon at the Bragg peak region in biological media. This paper reveals that the thermomechanical stress of the DNA molecule caused by the ion-induced shock wave becomes the dominant mechanism of complex DNA damage at the high-LET ion irradiation. Damage of the DNA molecule in water caused by a projectile ion induced shock wave is studied by means of reactive molecular dynamics simulations. Five projectile ions (carbon, oxygen, silicon, argon and iron) at the Bragg peak energies are considered. For the chosen segment of the DNA molecule and the collision geometry, the number of DNA strand breaks is evaluated for each projectile ion as a function of the bond dissociation energy and the distance from the ion's path to the DNA strands. Simulations reveal that argon and especially iron ions induce the breakage of multiple bonds in a DNA double convolution containing 20 DNA base pairs. The DNA damage produced in segments of such size leads to complex irreparable lesions in a cell. This makes the shock wave induced thermomechanical stress the dominant mechanism of complex DNA damage at the high-LET ion irradiation. A detailed theory for evaluating the DNA damage caused by ions at high-LET is formulated and integrated into the MSA formalism. The theoretical analysis reveals that a single ion hitting a cell nucleus at high-LET is sufficient to produce highly complex, lethal damages to a cell by the shock wave induced thermomechanical stress. Accounting for the shock wave induced thermomechanical mechanism of DNA damage provides an explanation for the ``overkill'' effect observed experimentally in the dependence of cell survival probabilities on the radiation dose delivered with iron ions. This important observation provides strong experimental evidence of the ion-induced shock wave effect and the related mechanism of radiation damage in cells.
\end{abstract}

\maketitle

\section{Introduction}

Experimental, theoretical and computational studies of radiation- and collision-induced processes with biomolecular systems are highly relevant nowadays in connection with the molecular-level assessment of biological damage induced by ionizing radiation
\cite{Gustavo_RadiatDamage, solov2016nanoscale, Surdutovich_2014_EPJD.68.353, surdutovich2019multiscale}.
The scientific interest in obtaining a deeper understanding of radiation damage is motivated by the development of radiotherapy with ion beams \cite{solov2016nanoscale, Amaldi_2005_RepProgPhys.68.1861, schardt2010heavy, Linz2012_IonBeams} and other applications of ions interacting with biological targets, e.g. radiation protection in space \cite{durante2011physical, Kronenberg2012_HealthPhys.103.556}.
Protons and carbon ions are currently used for cancer treatment, whereas the clinical implementation of other ions like helium and oxygen
has been discussed as the next step \cite{mein2019biophysical, tessonnier2018proton}.
Heavier ions can be found in galactic cosmic rays, where such elements as iron are present, being potentially damaging for humans during space missions \cite{durante2011physical}.

The mechanisms involved in radiation damage at the nanoscale and molecular level are still not entirely understood and are thus a subject of fundamental multidisciplinary research \cite{Gustavo_RadiatDamage, solov2016nanoscale, Surdutovich_2014_EPJD.68.353, surdutovich2019multiscale, Schuemann_SDD}.
The interaction of ion beams with biological materials has commonly been studied computationally by means of track-structure Monte Carlo simulations, which enable to follow the trajectory of each projectile, taking into account different physical interactions, such as elastic and inelastic scattering, electron transfer, nuclear fragmentation reactions, etc. \cite{Nikjoo_2016_RepProgPhys.79.116601}. Some Monte Carlo tools have recently included DNA models in the simulations of biodamage and the subsequent biological response (see \cite{Chatzipapas_2020_Cancers.12.799, Sakata_2020_SciRep.10.20788} and references therein).
Despite the wide use of the Monte Carlo approach for modeling ion propagation through biological media, it is unable to simulate the dynamics of the molecular medium in the vicinity of ion tracks, thus missing important physical phenomena.

It has been shown in recent years that a detailed physical understanding of the fundamental processes underlying radiation damage is indeed possible due to recent advances in the theoretical methods and experimental tools developed in atomic and molecular physics \cite{solov2016nanoscale}.
The phenomenon-based MultiScale Approach to the physics of radiation damage with ions (MSA) has been formulated and elaborated during the past decade (see \cite{solov2016nanoscale, solov2009physics, Surdutovich_2014_EPJD.68.353, surdutovich2019multiscale} and references therein).
This approach considers relevant physical, chemical and biological effects taking place on different scales in space, time and energy, and explores their manifestation in the biological damage.
The key phenomena and processes treated by the MSA are ion stopping in the medium, production of secondary electrons and free radicals as a result of ionization and excitation of the medium, transport of secondary electrons and reactive molecular species, the interaction of secondary particles with biomolecules, radiation chemistry, thermomechanical effects caused by nanoscale shock waves induced by ions, and the analysis of induced biodamage.
The important outcome of the MSA concerns the prediction of cell response to irradiation with ions on the basis of the assessment of complex DNA damage produced by a cascade of the aforementioned processes.
The MSA also demonstrated great success in predicting cell survival probabilities as a function of the radiation dose
in a wide range of the systems' parameters, including different cell types, ions with different values of linear energy transfer (LET), oxygenation level, as well as different cell repair conditions \cite{Surdutovich_2014_EPJD.68.353, surdutovich2019multiscale, verkhovtsev2016multiscale, verkhovtsev2019phenomenon}.

The important physical effect emphasized by the MSA concerns the manifestation of thermomechanical damage and related phenomena (e.g. transport of reactive secondary species) caused by nanoscale shock waves that are created by high-LET ions traversing biological medium \cite{surdutovich2010shock}.
The formation of ion-induced shock waves was predicted theoretically \cite{surdutovich2010shock} and studied
computationally in a series of subsequent papers \cite{Yakubovich_2011_AIP.1344.230, Yakubovich_2012_NIMB.279.135, surdutovich2013biodamage, bottlander2015effect, devera2016molecular, deVera_2018_EPJD.72.147, fraile2019first, ES_AVS_2019_EPJD.73.241, de2019role, friis_2020}.
This phenomenon arises due to the fact that ions can deposit a large amount of energy on the nanometer scale resulting in the significant heating up the medium in the localized vicinity of ion tracks.
The deposition of the energy lost by the ion into the medium occurs as a result of (i) production, transport and stopping of secondary electrons, and (ii) relaxation of the electronic excitation energy of the medium into its vibrational degrees of freedom through the electron--phonon coupling mechanism \cite{Gerchikov2000}.

The average kinetic energy of secondary electrons emitted in the vicinity of the Bragg peak is slightly below 40~eV \cite{Surdutovich_2014_EPJD.68.353}.
Electrons of such energy, experiencing both elastic and inelastic collisions, propagate up to $1-2$ nanometers away
from the ion's path within $\sim$50~fs before they become solvated electrons \cite{surdutovich2015transport}.
The radial distribution of secondary electrons emitted in the vicinity of the Bragg peak, obtained from the solution of the diffusion equation, is in agreement with the outcomes of track-structure Monte Carlo simulations \cite{Bug_2012_EPJD.66.291, Kyriakou_2016_JAP.119.194902}.

The energy lost by electrons in the processes of ionization and excitation of the medium is transferred to its heating (i.e. vibrational excitation of molecules) due to the electron--phonon interaction, enabling the electronic de-excitation of the molecules from the energy levels forbidden for other channels of de-excitation (such as autoionization, fragmentation or Auger processes).
As a result, the medium within the cylinder of the $\sim$1--2~nm radius surrounding the ion's path is heated up rapidly and the pressure inside this cylinder increases by several orders of magnitude (e.g. by a factor of 10$^3$ for a carbon ion at the Bragg peak \cite{Toulemonde2009_PRE}) compared to the pressure in the medium outside the cylinder.
High local temperature and pressure around the ion's path initiate a strong cylindrical explosion of the excited medium,
resulting in the formation of a shock wave \cite{surdutovich2010shock}.
Note that this effect has been yet unnoticed in the track-structure models based on the Monte Carlo approach although the classical theory of shock waves was established long ago \cite{LL_FluidMechanics_vol6, Zeldovich_ShockWaves}.
Note also that the ion-induced shock wave effect has been completely disregarded in the adaptation of the MSA formalism by other groups \cite{Kalospyros_2021_Molecules.26.840}.

The two possible mechanisms of DNA damage originating from the ion-induced shock wave have been suggested \cite{surdutovich2013biodamage, Surdutovich_2014_EPJD.68.353}. The shock wave may inflict damage by the thermomechanical stress and induce breakage of covalent bonds in the DNA molecule~\cite{Yakubovich_2011_AIP.1344.230, surdutovich2013biodamage, devera2016molecular, fraile2019first, bottlander2015effect, friis_2020}. Besides, the radial collective motion of the medium induced by the shock wave is instrumental in propagating the highly reactive molecular species, such as hydroxyl radicals and solvated electrons, to large radial distances (up to tens of nanometers) and preventing their recombination \cite{surdutovich2015transport, deVera_2018_EPJD.72.147}.

There are several strong evidences of the ion-induced shock wave effect.
First of all, as the shock wave spreads out, it becomes weaker and eventually turns into an acoustic wave at large distances from the ion's path. Acoustic waves coming from the Bragg peak region of ions' trajectories were detected experimentally \cite{Sun1993, Baily1992, Learned1979}.
Second, a similar phenomenon arising on the micrometer scale was observed during irradiation of micron-sized
water droplets with intense X-ray femtosecond pulses \cite{Stan2016_SW_xfel_NatPhys, Stan2016_SW_xfel_JPCL}.
Third, theoretical predictions for the radius and pressure on the shock wave front, based on the analytical solution of hydrodynamic equations \cite{surdutovich2010shock}, were supported by a series of molecular dynamics (MD) simulations \cite{Yakubovich_2011_AIP.1344.230, devera2016molecular, deVera_2018_EPJD.72.147, surdutovich2013biodamage, friis_2020}.
Finally, the inclusion of the shock wave effect in the multiscale scenario of biodamage with ions \cite{Surdutovich_2014_EPJD.68.353, surdutovich2019multiscale} has enabled to reproduce experimentally measured cell survival
probabilities and related radiobiological quantities such as oxygen enhancement ratio \cite{verkhovtsev2016multiscale, verkhovtsev2019phenomenon}.

In the earlier investigations \cite{Yakubovich_2011_AIP.1344.230, surdutovich2013biodamage, devera2016molecular}, the DNA damage by ion-induced shock waves was studied by means of classical MD simulations using non-reactive molecular mechanics force fields.
In those simulations the potential energy stored in a particular DNA bond was monitored in time as the bond length varied around its equilibrium distance \cite{surdutovich2013biodamage, devera2016molecular}.
When the potential energy of the bond exceeded a given threshold value, the bond was considered broken.
A more quantitative description of the phenomenon became possible by means of reactive MD simulations that permitted explicit simulation of covalent bond rupture and formation \cite{Sushko_2016_EPJD.70.12}. A recent study \cite{friis_2020} presented a detailed computational protocol for modeling the shock wave induced DNA damage by means of the reactive CHARMM (rCHARMM) force field \cite{Sushko_2016_EPJD.70.12}.

In this paper the thermomechanical stress of the DNA molecule caused by the ion-induced shock wave is systematically explored
by means of MD simulations with the rCHARMM force field following the aforementioned computational protocol \cite{friis_2020}.
Several projectile ions ranging from carbon to iron with different LET values corresponding to the Bragg peak region in liquid water are considered. The number of DNA strand breaks occurring in either one or both DNA strands is evaluated for each projectile ion as a function of the bond dissociation energy and the distance from the ion's path to the DNA strands. The simulations reveal that the shock wave induced thermomechanical stress by carbon and oxygen ions causes only a few isolated strand breaks within a DNA double twist containing 20 base pairs.
At higher LET values the thermomechanical stress induced by the shock wave becomes the dominant mechanism of DNA damage. This investigation reveals that argon and especially iron ions produce highly complex DNA damage consisting of multiple localized DNA strand breaks.

The quantitative information obtained from the performed MD simulations has been utilized to evaluate (by means of the MSA formalism) the survival probabilities of cells irradiated with ions.
It has been established that the shock wave affects the survival probabilities of cells irradiated with carbon ions mainly via the transport of reactive species away from the ion track. The shock wave induced by a single high-LET iron ion hitting a cell nucleus produces, in addition to the transport of reactive species, lethal damage to the cell due to the thermomechanical stress.
The accounting for this DNA damage mechanism within the MSA permits explaining the ``overkill'' effect, which arises when high-LET ions produce more biodamage than needed for the cell inactivation. A good agreement of the calculated cell survival probabilities with experimental data obtained for the cell irradiation with iron ions provides strong experimental evidence for the ion-induced shock wave effect.

\section{Methodology}
\label{sec:Methodology}

After setting up the all-atom model of a DNA molecule a series of reactive MD simulations have been performed while varying several parameters that characterize the interaction of an ion-induced shock wave with the target. The first part of this section describes the essentials of the computational protocol and introduces the different parameters used in the simulations. More details about this protocol are given in the recent study \cite{friis_2020}.
The second part of this section outlines the essentials of the MSA formalism regarding the evaluation of the number of DNA lesions produced by a projectile ion and the corresponding cell survival probability. The existing MSA formalism \cite{Surdutovich_2014_EPJD.68.353, surdutovich2019multiscale} is then extended to account for the shock-wave induced thermomechanical stress in the DNA damage caused by ion irradiation. It should be noted that VMD \cite{humphrey1996vmd} and MBN Studio \cite{sushko2019modeling} software have been used in the data analysis and visualization throughout the paper.

\subsection{Setting MD simulations of the DNA system}

In order to conduct simulations of DNA damage induced by the shock wave the system must first be constructed and undergo an extensive, multi-step equilibration process to correctly introduce the reactive rCHARMM force field \cite{Sushko_2016_EPJD.70.12} and ensure the system's stability before the simulation of the shock wave propagation. The methodology of designing and equilibrating the system was described in detail in our earlier study \cite{friis_2020}, and is therefore only briefly recapped below.

\begin{figure}[t!]
\centering
\includegraphics[width=0.8\textwidth]{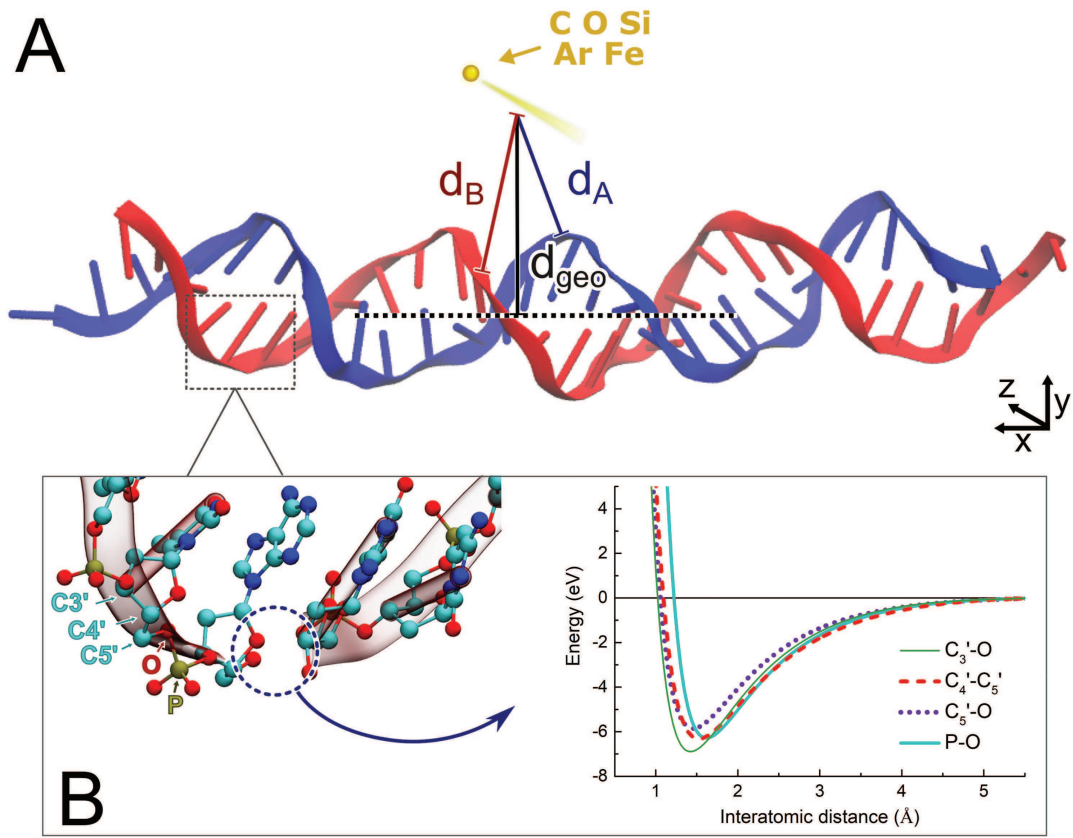}
\caption{Geometry of the DNA molecule and the studied parameters. Panel~A shows an ion (C, O, Si, Ar and Fe) propagating in close proximity to the DNA molecule consisting of 30 complementary base pairs. The ion track is oriented along the $z$-axis; the $x$-axis is oriented along one of the principal axis of inertia of the chosen DNA molecule, and the $y$-axis is along the line defining the shortest distance between the ion track and the selected principal axis of inertia. The collision parameter $d_{\text{geo}}$ is defined as the displacement of the ion's path along the $y$-axis with respect to the principal axis of inertia. The specific collision parameters $d_{\text{A}}$ and $d_{\text{B}}$ are defined as the shortest distances from the ion's path to DNA strand A and strand B, respectively. Panel~B illustrates C$_3^{\prime}$--O, C$_4^{\prime}$--C$_5^{\prime}$, C$_5^{\prime}$--O and P--O bonds in the DNA sugar-phosphate backbone and the corresponding potential energy curves obtained by means of DFT \cite{friis_2020}. Bond dissociation energy, $D_e$, defined as the depth of the associated potential energy well of the covalent bond is considered in the simulations as a variable parameter. The values of $D_e$ determined from the DFT calculations have been scaled by a factor of 2/3, 1/2, 1/3 and 1/6 (see main text for details).}
\label{fig. parameters}
\end{figure}

The investigated molecular system is created by joining together three short DNA segments (PDB-ID 309D \cite{qiu1997dna}) resulting in a double-stranded DNA molecule containing 30 complementary base pairs. The molecule is placed in a water box padding of 17 nm from the DNA in the $x$- and $y$-directions.
The coordinate system used in the simulations is illustrated in Fig.~\ref{fig. parameters}A. The $x$-axis of the coordinate system is oriented along the principal axis of inertia of the chosen DNA molecule with the largest moment of inertia at the initial time instance. The ion track is oriented along the $z$-axis. The $y$-axis is along the line defining the shortest distance between the ion track and the selected principal axis of inertia.
One sodium ion is placed for every phosphate group present in the DNA to ensure a neutral charge of the entire system, resulting in a system with a total of 1,010,994 atoms.
The whole system, including the DNA molecule and the water box, was equilibrated at 300~K temperature before the shock wave simulation.
After an initial equilibration in NAMD \cite{phillips2005scalable} with the standard CHARMM force field \cite{MacKerell_1998_JPCB.102.3586,mackerell2000development}, the system was transferred to the MBN Explorer software \cite{Solovyov_2012_MBNExplorer}, where the reactive rCHARMM force field \cite{Sushko_2016_EPJD.70.12} was used for further simulations.

rCHARMM is used to describe interatomic interactions in the C$_3^{\prime}$--O, C$_4^{\prime}$--C$_5^{\prime}$, C$_5^{\prime}$--O and P--O bonds in the DNA backbone, which connect the sugar ring of one nucleotide and the phosphate group of an adjacent nucleotide, see Fig.~\ref{fig. parameters}B.
Contrary to the standard CHARMM force field \cite{mackerell2000development} which employs a harmonic approximation for the description of covalent interactions (thereby limiting its applicability to small deformations of the molecular system), rCHARMM treats the bonded, angular and dihedral interactions differently \cite{Sushko_2016_EPJD.70.12}, thus permitting an accurate description of the molecular dissociation process in complex molecular systems. The radial dependence of the bonded interactions is described in rCHARMM by means of the Morse potential. The bonded interactions are set to zero for interatomic distances greater than a user-defined cutoff distance, beyond which the bond is considered broken and the molecular topology of the system is changed. Once a bond starts to break, the associated angular and dihedral interactions involving the indicated atoms weaken and eventually disappear when the distance between the atoms reaches a critical value \cite{Sushko_2016_EPJD.70.12}. Once all the associated bonded, angular and dihedral interactions go to zero, they are automatically removed from the molecular topology of the system. The atoms that initially formed the broken bond are then considered unbound, leading to the formation of atoms with dangling bonds.
Bond dissociation energies for the indicated bonds in the DNA strands and the cutoff distances for bond breakage/formation have been obtained by quantum chemistry calculations \cite{friis_2020}. Note that the C$_3^{\prime}$--C$_4^{\prime}$ bond was not parameterized by the rCHARMM force field because the dissociation energy of the C$_3^{\prime}$--C$_4^{\prime}$ bond is much higher (9.6~eV according to our DFT calculations) than the dissociation energies of the aforementioned bonds ($6.3 - 6.9$~eV) \cite{friis_2020}.

The DNA damage produced by the shock wave is systematically investigated for five different projectile ions propagating along the $z$-direction by varying the distance from the ion's path to the DNA strands and dissociation energies of the bonds in the DNA backbone. As described in detail in the following subsections, nine different values of the collision parameter $d_{\text{geo}}$ and five scaling factors for the bond dissociation energy $D_e$ have been considered for each projectile ion. For each simulation setup (i.e. for each ion type, values of $d_{\text{geo}}$ and $D_e$), two independent simulations of approximately 10~ps duration have been carried out. Hence, 450~independent simulations have been performed in total, and the total simulation time exceeded 2.5 million CPU hours.

\subsection{Setting up initial conditions for the shock wave simulation}

The shock wave is induced by an energetic ion propagating through the aqueous environment, where the ion loses its energy mainly
by electronic excitation and ionization of water molecules. For ions at the Bragg peak energies, ionization events result in the production of low-energy electrons (with the average kinetic energy of about 40~eV) which propagate radially on the nanometer scale away from the ion's path \cite{Surdutovich_2014_EPJD.68.353}.
Theoretical analysis of secondary electron transport revealed \cite{surdutovich2015transport} that sub-40~eV electrons lose most of their energy by ionizing and exciting molecules of the medium within approximately 1~nanometer from the ion's path in about 50~femtoseconds after the ion's passage through the medium. The electronic excitation energy of the medium is transferred into its vibrational degrees of freedom through the electron--phonon coupling mechanism \cite{Gerchikov2000}.
The relaxation of the energy deposited in close proximity to the ion track leads to a rapid increase of the temperature and the pressure of the medium around the ion track, resulting in the dynamical response of the medium and the formation of a cylindrical shock wave that propagates radially away from the ion track \cite{surdutovich2010shock}.
In a continuous medium this phenomenon is characterized by the so-called self-similar flow and the discontinuities of pressure and density of the medium at the wave front as follows from the analytical solution of a set of corresponding hydrodynamic and thermodynamic equations \cite{surdutovich2010shock}.

In the MD simulations, the energy lost by the propagating ion is deposited into the kinetic energy of water molecules located inside a ``hot'' cylinder of 1~nm radius around the ion's path. The radius of 1~nm is employed for all the ions considered in this study.
The equilibrium velocities of all atoms inside the ``hot'' cylinder are increased by a factor $\alpha$ such that the kinetic energy of these atoms reads as \cite{Yakubovich_2011_AIP.1344.230, Yakubovich_2012_NIMB.279.135, surdutovich2013biodamage}:
\begin{equation}
\sum_i^N \frac{1}{2} m_i \left( \alpha v_i \right)^2 = \frac{3 N k_{B} T}{2} + S_e \, l \ .
\label{eq:velocity_scaling}
\end{equation}
Here $S_e$ is the LET of the simulated ion, $l$ is the length of the simulation box in the $z$-direction (parallel to the ion's path), and $N$ is the total number of atoms within the ``hot'' cylinder. The first term on the right-hand side of Eq.~(\ref{eq:velocity_scaling}) is the kinetic energy of the 1-nm radius cylinder at the equilibrium temperature, $T = 300$~K, whereas the second term describes the energy loss by the ion as it propagates through the medium.
Note that the 1-nm radius for the energy deposition by low-energy secondary electrons was evaluated \cite{surdutovich2013biodamage} as the average distance at which secondary electrons lose most of their energy, according to the random walk approximation. The dispersion of the deposited energy due to more energetic secondary electrons (with the kinetic energy above 40~eV) and its impact on the dynamics of the ion-induced shock wave were addressed in the earlier study \cite{de2017radial}. It was demonstrated that, for ions at the Bragg peak, accounting for more energetic secondary electrons makes only a small correction to the results obtained for the uniform energy deposition within the cylinder of 1~nm radius around the ion's path. Since the present study is focused on the effects produced by ions at the Bragg peak region, the utilized ``hot'' cylinder model captures all the relevant phenomena correctly.

\subsection{Parameters for the reactive MD simulations}
\label{sec:simulation_parameters}

DNA damage caused by the ion-induced shock wave is simulated for five different ions varying the distance from the ion's path to the DNA molecule and dissociation energies of bonds in the DNA backbone, see Fig.~\ref{fig. parameters}.
The choice of the specific parameters is explained and justified below.

\subsubsection{Distance from the ion's path to DNA strands}

In the simulations each ion propagated along the $z$-axis orthogonal to the principal axis of inertia of the DNA molecule with the largest moment of inertia at the initial time instance, below called simply the principle axis of inertia.
The collision parameter $d_{\text{geo}}$, defined as the displacement of the ion's path with respect to the principal axis of inertia, varied from 0 to 12~\AA~with an increment step of 3~\AA. The ion's path was considered at the positive and the negative directions along the $y$-axis resulting in the positive and negative values of $d_{\text{geo}}$.
To account for the orientation of DNA strands with respect to the ion's path, the collision parameter was related to the shortest distance to strand A, $d_{\text{A}}$, and the shortest distance to the strand B, $d_{\text{B}}$.
As such, an increase of the displacement $d_{\text{geo}}$ could result simultaneously in an increased distance to one strand and a decreased distance to the other strand. Geometry of the system is illustrated in Fig.~\ref{fig. parameters}A, whereas the values of the considered parameters $d_{\text{geo}}$, $d_{\text{A}}$ and $d_{\text{B}}$ are listed in Table~\ref{tab:distances}.

\begin{table}[tb!]
	\centering
	\begin{tabular}{c|p{0.9cm} p{0.9cm} p{0.9cm} p{0.9cm} p{0.9cm} p{0.9cm} p{0.9cm} p{0.9cm} p{0.9cm}}
    \hline
		$d_{\text{geo}}$ (\AA) & 0   & 3   & $-3$ & 6   & $-6$ & 9   & $-9$ & 12  & $-12$ \\ \hline
		$d_{\text{A}}$ (\AA)   & 3.9 & 1.5 & 5.9  & 0.4 & 5.4  & 2.6 & 5.4  & 5.4 & 6.3   \\
		$d_{\text{B}}$ (\AA)   & 2.5 & 4.5 & 1.1  & 5.5 & 0.4  & 6.7 & 1.0  & 8.7 & 2.3   \\ \hline
	\end{tabular}
	\caption{Collision parameter values used in the simulations. The table summarizes the displacement of the ion's path along the $y$-axis with respect to the principal axis of inertia of the DNA molecule, $d_{\text{geo}}$, and the respective shortest distances to strand A, $d_{\text{A}}$, and strand B, $d_{\text{B}}$.}
	\label{tab:distances}
\end{table}

\subsubsection{Dissociation energy of covalent bonds in the DNA backbone}

The number of DNA strand breaks induced by the shock wave impact may depend on the energy required to break covalent bonds. The typical dissociation energy of covalent bonds in the DNA backbone varies from about 3 to 6~eV \cite{range2004structure}.
The deposition of such an amount of energy into a given bond would most likely lead to its instantaneous rupture. However, it has also been established that the threshold energy for bond dissociation can be several times smaller due to the presence of solvated electrons in the molecular medium surrounding the DNA. For instance, it was shown \cite{smyth2012excess,li2003density} that attachment of a solvated electron to a DNA molecule decreases the dissociation energy of covalent bonds in the backbone down to $\sim$1~eV and leads predominantly to cleavage of a phosphodiester bond.
In the present study the bond dissociation energy $D_e$ is considered as a variable parameter to account for different possible scenarios that happen on the femto- to sub-picosecond time scales preceding the shock wave formation. A detailed analysis of the DNA damage events created by secondary electrons on the indicated time scales is beyond the scope of this study.
Dissociation energies for several bonds along the DNA backbone, shown in Fig.~\ref{fig. parameters}B, were determined from density functional theory (DFT) calculations \cite{friis_2020}. The obtained values are scaled by a factor of 2/3, 1/2, 1/3 and 1/6
to account for the weakening of the bonds, which may happen e.g. upon the attachment of solvated electrons. The resulting bond dissociation energies thus vary from about 1~eV to 6~eV; this range corresponds to the range of values reported in \cite{range2004structure,smyth2012excess,li2003density}.

\subsubsection{Different projectile ions}

The number of shock wave induced DNA strand breaks also depends on the type of ions irradiating the biological target.
Carbon ions are presently used as radiation modality in ion-beam cancer treatments \cite{solov2016nanoscale, Amaldi_2005_RepProgPhys.68.1861, schardt2010heavy, Linz2012_IonBeams}, whereas the interaction with heavier ions (up to iron) is particularly relevant for the radiation protection of astronauts during manned space missions \cite{durante2011physical}.
In the present study shock waves induced by five different projectile ions (C$^{6+}$, O$^{8+}$, Si$^{14+}$, Ar$^{18+}$ and Fe$^{26+}$) with energies corresponding to the Bragg peak region in liquid water are analyzed.

The LET $S_e$ as a function of projectile's kinetic energy $E$ has been calculated using the analytical MSA model described in detail in earlier studies~\cite{Surdutovich_2009_EPJD.51.63, Surdutovich_2014_EPJD.68.353, Scifoni_2010_PRE.81.021903}.
The model is based upon the Rudd's formalism \cite{Rudd1992} which is extended to account for relativistic corrections and an effective charge of the projectile that arises when a bare ion picks off electrons while propagating through a medium.
The dependence of LET on $E$ then reads as:
\begin{equation}
S_e(E) = - \frac{dE}{dx} = n \sum_i  \int\limits_0^{\infty} (W + I_i) \frac{d\sigma_i}{dW} dW \ \label{eq. LET},
\end{equation}
where $n$ is the number density of water molecules in the medium and $W$ is the kinetic energy of ejected electrons.
The sum on the right-hand side is taken over all electron shells of the water molecule with $I_i$ being the binding energy of the $i$th electron shell and $d\sigma_i/dW$ the partial single differential ionization cross section of the corresponding shell.
Parameters of the analytical MSA model for liquid water are taken from \cite{Dingfelder_2000_RPC.59.255}.
Although these parameters were originally derived for proton--water interactions, they are also applicable for evaluating the LET of heavier ions, as illustrated below.

Solid lines in Figure~\ref{fig:LET}A show the $S_e(E)$ dependence for C$^{6+}$, O$^{8+}$, Si$^{14+}$, Ar$^{18+}$ and Fe$^{26+}$ ions calculated using Eq.~(\ref{eq. LET}). The results are compared with the values compiled in the ICRU73 report \cite{ICRU_Report_73} (open symbols) and the results of Monte Carlo simulations \cite{Francis_2011_PMB.57.209} performed using the Geant4-DNA software package \cite{Bernal_2015_PhysMed.31.861} (closed symbols).
Figure~\ref{fig:LET}B shows a detailed comparison of the calculated LET for carbon ions with the results of recent experiments \cite{Rahm_2014_PMB.59.3683} (open triangles) as well as with other theoretical calculations performed using the popular SRIM \cite{SRIM_Ziegler_2010_NIMB} and CasP \cite{CasP_Schiwietz_2011_PRA.84.052703} codes.
The $S_e(E)$ dependence calculated using Eq.~(\ref{eq. LET}) (thick solid line) gives the best agreement with the experimental data for carbon ions \cite{Rahm_2014_PMB.59.3683} in terms of both the position of the Bragg peak and its magnitude.
Reportedly, there is no experimentally measured $S_e(E)$ dependence for ions heavier than carbon, and the comparison can only be made with the results of other calculations or Monte Carlo simulations.
As shown in Fig.~\ref{fig:LET}, there is some deviation (about $10-15$\% in the Bragg peak region) between the results obtained with different theoretical methods. The results of the present analysis fit nicely into this range of values.
Table~\ref{tab: LET_BP} lists the values of LET for each ion at the Bragg peak in liquid water and provides the respective ion's kinetic energy. The values from Table~\ref{tab: LET_BP} have been used in Eq.~(\ref{eq:velocity_scaling}) to scale the velocities of atoms of the medium located within the ``hot'' cylinder for the MD simulations of shock wave propagation.

\begin{figure}[tb!]
\centering
\includegraphics[width=0.85\textwidth]{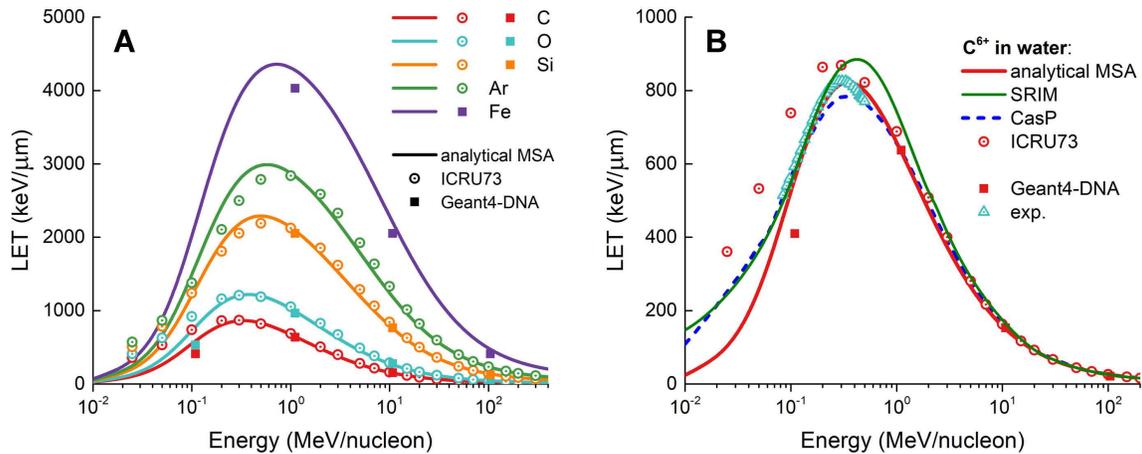}
\caption{\textbf{A}: The LET for C$^{6+}$, O$^{8+}$, Si$^{14+}$, Ar$^{18+}$ and Fe$^{26+}$ ions as a function of ion's kinetic energy, calculated using the analytical MSA model \cite{Surdutovich_2014_EPJD.68.353} (solid lines). The results are compared with the values compiled in the ICRU73 report \cite{ICRU_Report_73} (open symbols) and the results of Monte Carlo simulations \cite{Francis_2011_PMB.57.209} performed using the Geant4-DNA software package \cite{Bernal_2015_PhysMed.31.861} (closed symbols).
\textbf{B}: Comparison of the calculated LET for carbon ions (thick solid line) with experimental measurements \cite{Rahm_2014_PMB.59.3683} (open triangles) and other theoretical calculations performed using the widely-used SRIM \cite{SRIM_Ziegler_2010_NIMB} (thin solid line) and CasP \cite{CasP_Schiwietz_2011_PRA.84.052703} (dashed line) codes.}
\label{fig:LET}
\end{figure}

\begin{table}[t!]
	\centering
	\begin{tabular}{c|c|c}
        \hline
		Ion        &  $S_e$ (keV/$\mu$m) & $E$ (MeV/u)  \\
		\hline
		C$^{6+}$   &  830  & 0.35  \\
		O$^{8+}$   & 1220  & 0.38  \\
		Si$^{14+}$ & 2200  & 0.54  \\
		Ar$^{18+}$ & 2890  & 0.63  \\
		Fe$^{26+}$ & 4230  & 0.80  \\
        \hline
	\end{tabular}
	\caption{Linear energy transfer, $S_e$, at the Bragg peak region for different ions considered in this study and the corresponding ion's kinetic energy, $E$.}
	\label{tab: LET_BP}
\end{table}

\subsection{Shock wave propagation in pure water}

In order to quantify the impact of the shock wave on the transport of reactive molecular species, additional MD simulations of a shock wave propagating in pure water have been performed following the computational protocol described in \cite{friis_2020}.
The water box dimensions were set to 49.5~nm $\times$ 49.5~nm $\times$ 8~nm.
No DNA molecule or neutralizing ions were included, so that the shock wave propagated in liquid water. Simulations for the shock wave induced by silicon, argon and iron ions were carried out for $\sim$10~ps, while the simulations for the lighter (carbon and oxygen) ions were performed for $\sim$30~ps.

\section{Evaluation of the number of ion-induced DNA lesions and cell survival probability}
\label{sec:MSA_methodology}

The MSA formalism has been developed to describe survival probabilities of cells irradiated with ion beams on the basis of detailed physical understanding of the fundamental processes underlying radiation damage by ions \cite{Surdutovich_2014_EPJD.68.353, surdutovich2019multiscale, verkhovtsev2016multiscale}. As described above, all the relevant physical, chemical and biological processes and phenomena are interlinked within the MSA into a unified multiscale scenario of ion-induced biodamage.
A comprehensive description of the MSA formalism is presented in \cite{solov2016nanoscale, Surdutovich_2014_EPJD.68.353, surdutovich2019multiscale}. This section outlines the formalism for evaluating the number of lesions of the DNA molecule produced upon its irradiation with ions and the corresponding cell survival probabilities. The case study is focused on the projectile ions in the vicinity of the Bragg peak. The previously developed formalism \cite{solov2016nanoscale, Surdutovich_2014_EPJD.68.353, surdutovich2019multiscale} is extended towards accounting for the DNA lesions produced by the thermomechanical stress imposed on the DNA molecule by the propagating shock wave.

The starting point for this theory is the calculation of ${\cal N}(r, S_e)$  -- the total average number of simple lesions, i.e. single-strand breaks (SSBs), produced in a DNA double convolution (a DNA double twist) located at distance $r$ from the ion's path. This number depends on the ion's type and its LET $S_e$. According to the MSA analysis \cite{solov2016nanoscale, Surdutovich_2014_EPJD.68.353}, ${\cal N}(r, S_e)$ is equal to
\begin{equation}
{\cal N}(r, S_e) = {\cal N}_{{\rm e}}(r, S_e) + {\cal N}_{{\rm r}}(r, S_e) + {\cal N}_{{\rm SW}}(r, S_e) \ .
\label{eq:MSA_numlesions}
\end{equation}
Here
\begin{equation}
{\cal N}_{{\rm e}}(r, S_e) = \Gamma_{{\rm e}} \, F_{{\rm e}}(r, S_e)
\label{eq:MSA_numlesions_Ne}
\end{equation}
is the number of simple lesions produced by secondary electrons.
The function $F_{{\rm e}}(r, S_e)$ is the number of electrons incident on the DNA segment located at distance $r$ from the ion's path.
The quantity $\Gamma_{{\rm e}}$ is the average probability of producing a SSB per electron hit. For ions in the Bragg peak region the probability $\Gamma_{{\rm e}}$ does not depend on $S_e$ since the average kinetic energy of produced secondary electrons is about 40~eV for different ions with different LET values \cite{Surdutovich_2014_EPJD.68.353}.
For ions outside the Bragg peak region the dependence of $\Gamma_{{\rm e}}$ on $S_e$ should arise as the kinetic energy of produced $\delta$-electrons is LET-dependent. The analysis of this regime goes beyond the scope of the present study.

The second term on the right-hand side of Eq.~(\ref{eq:MSA_numlesions}),
\begin{equation}
{\cal N}_{{\rm r}}(r, S_e) = {\cal N}_{{\rm r},0}(S_e) \, \theta(R_{{\rm r}}(S_e) - r) \ ,
\label{eq:MSA_numlesions_Nr}
\end{equation}
is the number of lesions produced by free radicals that are uniformly spread over the distances $r < R_{{\rm r}}(S_e)$ defined by the radius of shock wave propagation. $\theta(x)$ on the right-hand side of Eq.~(\ref{eq:MSA_numlesions_Nr}) is the Heaviside step function.
A linear dependence $R_{{\rm r}} \propto S_e$ was explored in the earlier study \cite{surdutovich2017ion}, and a conservative estimate
$R_{{\rm r}} \approx 10$~nm was derived for carbon ions in the Bragg peak region \cite{Surdutovich_2014_EPJD.68.353}.
In the present paper the $R_{{\rm r}}$ value for carbon ions is evaluated more precisely on the basis of MD simulations, and the $R_{{\rm r}}$ values for heavier ions are estimated according to the $R_{{\rm r}} \propto S_e$ dependence from the analysis of the pressure at the shock wave front, see Sect.~\ref{sec:SW_propagation}.

The value ${\cal N}_{{\rm r},0}(S_e)$ depends on the number of formed free radicals, which in turn is proportional to the number of generated secondary electrons and hence proportional to LET.
${\cal N}_{{\rm r},0}(S_e)$ depends also on the degree of oxygenation of the medium since the concentration of oxygen dissolved in the medium affects the number of formed radicals and, consequently, the creation of DNA lesions.
For carbon ions at the Bragg peak, the value ${\cal N}_{{\rm r},0} = 0.08$ for the environment with the normal concentration of oxygen was derived earlier \cite{verkhovtsev2016multiscale} from the comparison of the experimental results \cite{dang2011heavy} for plasmid DNA, dissolved in pure water and in a scavenger-rich solution, and irradiated with carbon ions at the Bragg peak region.
A number of cell survival experiments performed at hypoxic conditions were reproduced with the twice smaller value of ${\cal N}_{{\rm r},0} = 0.04$ \cite{solov2016nanoscale, verkhovtsev2016multiscale}.

The third term on the right-hand side of Eq.~(\ref{eq:MSA_numlesions}), ${\cal N}_{{\rm SW}}(r, S_e)$, is the number of DNA lesions produced by the thermomechanical stress imposed on the DNA molecule by the propagating shock wave.

The creation of DNA lesions by secondary electrons, free radicals and the shock wave are statistically independent events taking place at different time scales after the ion passage \cite{solov2016nanoscale, Surdutovich_2014_EPJD.68.353}. 
Therefore, the total average number of simple lesions in a DNA double twist, ${\cal N}(r, S_e)$, is a cumulative quantity derived by integrating all the events over time. ${\cal N}_{{\rm e}}(r, S_e)$ and ${\cal N}_{{\rm r}}(r, S_e)$ were worked out earlier within the MSA \cite{solov2016nanoscale, Surdutovich_2014_EPJD.68.353, verkhovtsev2016multiscale, surdutovich2015transport, surdutovich2019multiscale}, whereas ${\cal N}_{{\rm SW}}(r, S_e)$ is quantified in the present study by means of MD simulations.

Knowing ${\cal N}(r, S_e)$ at a given distance $r$ and for a given ion's LET $S_e$, one can use the Poisson statistics to calculate probabilities of different independent events. The probability to produce $k$ lesions in a DNA double twist placed at a distance $r$ from the ion track is equal to
\begin{equation}
{\cal P}_k(r, S_e) = \frac{ {\cal N}^k(r, S_e) }{k!} \, e^{ - {\cal N}(r, S_e)} \ .
\label{eq:MSA_Poisson_Pk}
\end{equation}
A lethal DNA lesion is defined within the MSA framework as one double-strand break (DSB) plus at least two additional single lesions occurring within a DNA double twist \cite{Surdutovich_2014_EPJD.68.353}. This definition relies on earlier findings \cite{schipler2013dna, Mavragani_2017_Cancers, Nickoloff2020_Genes} that complex DNA damage is irreparable for a cell if the damage occurs in a localized DNA segment, which typically consists of two helical turns containing 20~base pairs.

Lesions within the DNA double twist may occur on one DNA strand or be present on both strands.
As shown in Fig.~\ref{fig. parameters} each nucleotide in the DNA molecule has four vulnerable covalent bonds in the sugar-phosphate backbone. Therefore the number of such covalent bonds in one strand in a DNA double twist is equal to 80, and the total number of such bonds in both strands in the DNA double twist is $2n = 160$.

The total number of events $N_{\nu}$ for $\nu = 0, 1, ..., 2n$ lesions occurring within the DNA double twist is equal to the number of combinations for $\nu$ choices taken out of $2n$ places:
\begin{equation}
N_{\nu} = C_{2n}^{\nu} \equiv \frac{(2n)!}{(2n-\nu)! \, \nu!}  \ , \qquad \nu = 0, 1, ..., 2n \ .
\label{eq:N_nu-1}
\end{equation}
On the other hand, the number of independent events of $k$ lesions occurring in a single DNA strand of the length $n$ is equal to $C_n^k$.
Therefore, $N_{\nu}$ can be calculated as follows:
\begin{equation}
N_{\nu} = \sum_{k=0}^{\nu} C_n^{k} \, C_n^{\nu-k} \equiv \frac{(2n)!}{(2n-\nu)! \, \nu!}  \ , \qquad \nu = 0, 1, ..., n \ .
\label{eq:N_nu-2}
\end{equation}
This relationship is well-known, see e.g. Eq.~(0.156) in \cite{Gradshteyn_Ryzhik}.
In the case $\nu = n+1, n+2, ..., 2n$ the number of events $N_{\nu}$ is equal to
\begin{equation}
N_{\nu} = \sum_{k=\nu-n}^{n} C_n^{k} \, C_n^{\nu-k} \ , \qquad \nu = n+1, n+2, ..., 2n \ .
\label{eq:N_nu-3}
\end{equation}
Here $k = \nu-n$ is the minimum number of lesions on a DNA strand if the other strand within the DNA double twist possesses $n$ lesions. Substituting $k = k^{\prime} + \nu - n$ in Eq.~(\ref{eq:N_nu-3}), one derives
\begin{equation}
N_{\nu} = \sum_{k^{\prime} = 0}^{2n-\nu} C_n^{k^{\prime}+\nu-n} \, C_n^{n-k^{\prime}} \ .
\end{equation}
Noting that $C_n^{n-k} = C_n^k$ and using Eq.~(0.156(2)) from \cite{Gradshteyn_Ryzhik}, one derives the same relationship as in Eq.~(\ref{eq:N_nu-2}), but now valid for $\nu = n+1, n+2, ..., 2n$. This proves that counting of the lesion events occurring on both DNA strands leads to the same result for $N_{\nu}$ as given by Eq.~(\ref{eq:N_nu-1}).

Similarly, the number of events $N_{\nu}^{(1)}$ of $\nu$ lesions being all located on one strand within the DNA double twist can be calculated as
\begin{equation}
N_{\nu}^{(1)} = \left\{
\begin{array}{l l}
1         & \ , \quad \nu = 0
\\
2 \, C_n^{\nu} \equiv \displaystyle{ 2 \, \frac{n!}{(n-\nu)! \, \nu!} }         & \ , \quad \nu = 1, 2, ..., n
\\
0           & \ , \quad \nu = n+1, n+2, ..., 2n \ .
\end{array} \right.
\label{eq:N_nu1}
\end{equation}
The number of events $N_{0}^{(1)}$ corresponding to the absence of lesions ($\nu = 0$) is naturally equal to one. For $\nu = 1, 2, ..., n$ lesions the factor 2 accounts for the two strands within the DNA double twist. The larger number of lesions ($\nu = n+1, n+2, ... 2n$) will necessarily occur on both DNA strands, thus the corresponding numbers $N_{\nu}$ are equal to zero.

One can also calculate the number of events $N_{\nu}^{(2)}$ when $\nu$ lesions result in at least one DSB within the DNA double twist:
\begin{equation}
N_{\nu}^{(2)} = \left\{
\begin{array}{l l}
0          & \ , \quad \nu = 0, 1
\\
\displaystyle{ \sum_{k=1}^{\nu-1} C_n^{k} \, C_n^{\nu-k} }       & \ , \quad \nu = 2, 3, ..., n
\\
\displaystyle{ \frac{(2n)!}{(2n-\nu)! \, \nu!} } &  \ , \quad \nu = n+1, n+2, ..., 2n \ .
\end{array} \right.
\label{eq:N_nu2}
\end{equation}

The numbers $N_{\nu}$, $N_{\nu}^{(1)}$ and $N_{\nu}^{(2)}$ from Eqs.~(\ref{eq:N_nu-1}), (\ref{eq:N_nu1}) and (\ref{eq:N_nu2}) obey the obvious relationship
\begin{equation}
N_{\nu} = N_{\nu}^{(1)} + N_{\nu}^{(2)} \ .
\label{eq:nu1+nu2}
\end{equation}

Knowing $N_{\nu}^{(1)}$ and the total number of events for $\nu$ lesions, $N_{\nu}$, one derives the probability ${\cal P}_{\nu}^{(1)}$ to create $\nu$ SSBs located on one DNA strand within the double twist:
\begin{equation}
{\cal P}_{\nu}^{(1)} = \frac{ N_{\nu}^{(1)} }{ N_{\nu} } \ , \qquad \nu = 0, 1, ..., 2n   \ .
\label{eq:Prob_nu1-1}
\end{equation}
Substituting here $N_{\nu}$ and $N_{\nu}^{(1)}$ from Eqs.~(\ref{eq:N_nu-1}) and (\ref{eq:N_nu1}) respectively, one derives
\begin{equation}
{\cal P}_{\nu}^{(1)} = \left\{
\begin{array}{l l}
1         & \ , \quad \nu = 0
\\
\displaystyle{ 2 \, \frac{ n! }{(n - \nu)!} \, \frac{ (2n - \nu)! }{(2n)!}  }       & \ , \quad \nu = 1, 2, ..., n
\\
0           & \ , \quad \nu = n+1, n+2, ..., 2n \ .
\end{array} \right.
\label{eq:Prob_nu1-2}
\end{equation}

Analogously, the probability ${\cal P}_{\nu}^{(2)}$ that $\nu$ lesions result in at least one DSB within the DNA double twist reads as
\begin{equation}
{\cal P}_{\nu}^{(2)} = \frac{ N_{\nu}^{(2)} }{ N_{\nu} } \ , \qquad \nu = 0, 1, ..., 2n   \ .
\label{eq:Prob_nu2-1}
\end{equation}
Substituting $N_{\nu}^{(2)}$ from Eq.~(\ref{eq:N_nu2}) and using Eqs.~(\ref{eq:nu1+nu2})--(\ref{eq:Prob_nu1-2}) one derives
\begin{equation}
{\cal P}_{\nu}^{(2)} = \left\{
\begin{array}{l l}
0         & \ , \quad \nu = 0,1
\\
\displaystyle{ 1 - {\cal P}_{\nu}^{(1)} \equiv  1 - 2 \, \frac{ n! }{(n - \nu)!} \, \frac{ (2n - \nu)! }{(2n)!}  }     & \ , \quad \nu = 2, 3, ..., n
\\
1         & \ , \quad \nu = n+1, n+2, ..., 2n \ .
\end{array} \right.
\label{eq:Prob_nu2-2}
\end{equation}

Now following the above introduced criterion for a lethal DNA lesion, one can derive the probability of such event as follows:
\begin{eqnarray}
{\cal P}_{l}(r, S_e) &=&
\lambda \sum_{\nu = 3}^{\nu_{{\rm max}}}
{\cal P}_{\nu}^{(1)} \, \frac{ {\cal N}^{\nu}(r, S_e) }{\nu!} \, e^{ -{\cal N}(r, S_e) }
\nonumber \\ &+&
\lambda \, {\cal P}_{3}^{(2)} \, \frac{ {\cal N}^3(r, S_e) }{3!} \, e^{ -{\cal N}(r, S_e) }
+
\sum_{\nu = 4}^{\nu_{{\rm max}}}
{\cal P}_{\nu}^{(2)} \, \frac{ {\cal N}^{\nu}(r, S_e) }{\nu!} \, e^{ -{\cal N}(r, S_e) } \ .
\label{eq:MSA_prob_lethal}
\end{eqnarray}
Here $\lambda$ is the probability that a SSB can be converted to a DSB and $\nu_{{\rm max}} = 2n$.
Accounting for $\lambda$ relies on the experimental findings \cite{Huels2003single, sanche2005low} that the DSBs caused by low-energy electrons with energies higher than $\sim$5~eV happen in one hit. In that case the subsequent break in the second DNA strand occurs due to the action of debris generated by the first SSB.
Following \cite{Huels2003single, sanche2005low} $\lambda$ is set equal to 0.15 within the MSA framework \cite{Surdutovich_2014_EPJD.68.353}.

The first term on the right-hand side of Eq.~(\ref{eq:MSA_prob_lethal}) describes the sum of probabilities to have all $\nu$ ($\nu = 3, 4, ..., 2n$) lesions on one DNA strand with the subsequent conversion of one SSB into a DSB.
The second term is the probability of three lesions with at least one DSB among them and the subsequent conversion of one SSB into a DSB, i.e. creating two DSBs.
The third term is the sum of probabilities of $\nu$ lesions ($\nu = 4, 5, ..., 2n$) with creation of at least one DSB.

After simple algebraic transformations Eq.~(\ref{eq:MSA_prob_lethal}) can be rewritten in the form:
\begin{equation}
{\cal P}_{l}(r, S_e)
=
\lambda \sum_{\nu = 3}^{\nu_{{\rm max}}}
\frac{ {\cal N}^{\nu}(r, S_e) }{\nu!} \, e^{-{\cal N}(r, S_e)}
+
(1 - \lambda) \sum_{\nu = 4}^{\nu_{{\rm max}}}
{\cal P}_{\nu}^{(2)} \, \frac{ {\cal N}^{\nu}(r, S_e) }{\nu!} \, e^{-{\cal N}(r, S_e)} \ ,
\label{eq:MSA_prob_lethal_final}
\end{equation}
with ${\cal P}_{\nu}^{(2)}$ defined above in Eq.~(\ref{eq:Prob_nu2-2}).

\begin{figure*}[t!]
\centering
\includegraphics[width=0.85\textwidth]{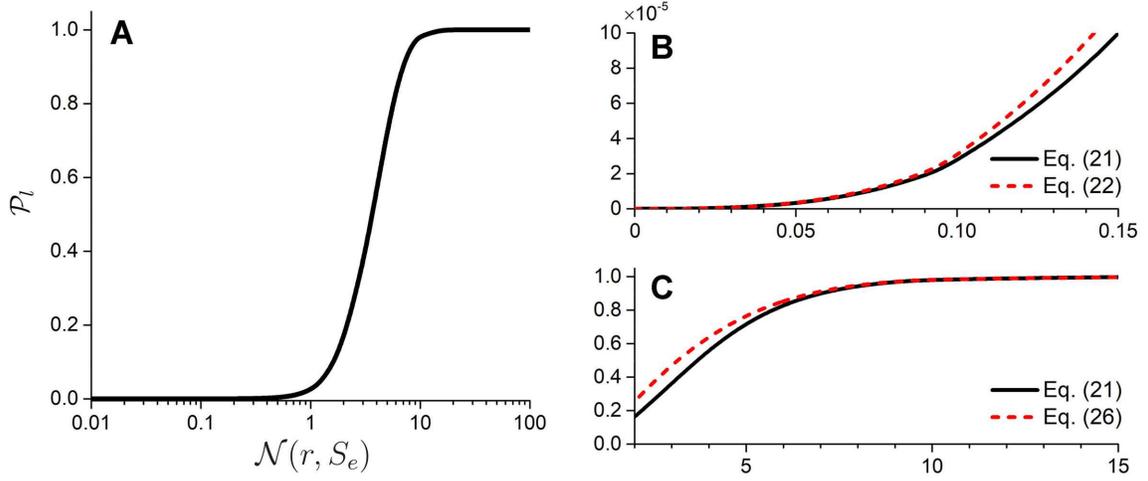}
\caption{The dependence of the probability for a lethal DNA lesion, ${\cal P}_{l}$, on the average number of simple lesions within the DNA double twist, ${\cal N}(r, S_e)$. \textbf{A:} The dependence calculated according to Eq.~(\ref{eq:MSA_prob_lethal_final+SW}). In panels (\textbf{B}) and (\textbf{C}) this dependence is compared with the limiting case of ${\cal N}(r, S_e) \lesssim \lambda \ll 1$ calculated according to Eq.~(\ref{eq:MSA_prob_lethal_final+SW-2a}) and with the limiting case of $1 \ll {\cal N}(r, S_e) \ll \nu_{{\rm max}}$ calculated according to Eq.~(\ref{eq:MSA_prob_lethal_final+SW-3}), respectively.}
\label{fig:prob_lethal_calN}
\end{figure*}

Let us introduce the upper incomplete gamma function \cite{Gradshteyn_Ryzhik}
\begin{equation}
\Gamma(n+1, x) = n! \, e^{-x} \sum_{m=0}^{n} \frac{x^m}{m!}  \ , \qquad n = 0, 1, 2, ...
\label{eq:gamma_function}
\end{equation}
and rewrite Eq.~(\ref{eq:MSA_prob_lethal_final}) in the form
\begin{eqnarray}
{\cal P}_{l}(r, S_e) &=&
\lambda \, \left[ \frac{\Gamma(\nu_{{\rm max}}+1,{\cal N}(r, S_e))}{\nu_{{\rm max}}!}  -
e^{-{\cal N}(r, S_e)} \left(  1 + {\cal N}(r, S_e) + \frac12 {\cal N}^2(r, S_e) \right)
\right] \nonumber \\
&+&
(1 - \lambda)
\sum_{\nu = 4}^{\nu_{{\rm max}}}
{\cal P}_{\nu}^{(2)} \, \frac{ {\cal N}^{\nu}(r, S_e) }{\nu!} \, e^{-{\cal N}(r, S_e)} \ .
\label{eq:MSA_prob_lethal_final+SW}
\end{eqnarray}
The dependence of ${\cal P}_{l}$ on ${\cal N}(r, S_e)$ calculated according to Eq.~(\ref{eq:MSA_prob_lethal_final+SW}) is shown in Fig.~\ref{fig:prob_lethal_calN}A.

At small LET values when the number of lesions in a DNA double twist ${\cal N}(r, S_e) \lesssim \lambda \ll 1$, the probability of lethal events ${\cal P}_{l}(r, S_e)$ is simplified to
\begin{equation}
{\cal P}_{l}(r, S_e)
\simeq \lambda \, \frac{{\cal N}^3(r, S_e)}{3!} + \frac{7}{8} \, \frac{ {\cal N}^{4}(r, S_e) }{4!} \ .
\label{eq:MSA_prob_lethal_final+SW-2a}
\end{equation}

If the characteristic number of lesions is much smaller than the total number of bonds in the DNA double twist, $\nu \ll \nu_{{\rm max}} = 2n$, the probability $P_{\nu}^{(2)}$, Eq.~(\ref{eq:Prob_nu2-2}), is reduced to
\begin{equation}
{\cal P}_{\nu}^{(2)} \simeq 1 - \frac{1}{2^{\nu-1}} \ .
\end{equation}
Then one derives from Eq.~(\ref{eq:MSA_prob_lethal_final+SW}) the following expression
\begin{eqnarray}
{\cal P}_{l}(r, S_e)
&\simeq&
\lambda \, \left[ \frac{\Gamma(\nu_{{\rm max}} + 1,{\cal N}(r, S_e))}{\nu_{{\rm max}}!}  -
e^{-{\cal N}(r, S_e)} \left(  1 + {\cal N}(r, S_e) + \frac12 {\cal N}^2(r, S_e) \right) \right] \nonumber \\
&+&
(1 - \lambda)
\sum_{\nu = 4}^{\nu_{{\rm max}}}
\left( 1 - \frac{1}{2^{\nu-1}} \right) \, \frac{ {\cal N}^{\nu}(r, S_e) }{\nu!} \, e^{-{\cal N}(r, S_e)} \ .
\label{eq:MSA_prob_lethal_final+SW-2b}
\end{eqnarray}

Now let us consider the region ${\cal N}(r, S_e) \gg 1$.
Using the definition of the function $\Gamma(n+1, x)$, Eq.~(\ref{eq:gamma_function}), the fact that
\begin{equation}
\frac{\Gamma(\nu_{{\rm max}} + 1,{\cal N}(r, S_e))}{\nu_{{\rm max}}!} \simeq 1
\end{equation}
at $1 \ll {\cal N}(r, S_e) \ll \nu_{{\rm max}}$, and keeping only the leading terms in Eq.~(\ref{eq:MSA_prob_lethal_final+SW-2b}), one derives
\begin{equation}
{\cal P}_{l}(r, S_e)
\simeq
1  - 2(1 - \lambda) \, e^{-\frac{{\cal N}(r, S_e)}{2}} - \frac{3}{24} (1 - \lambda) \, e^{-{\cal N}(r, S_e)} \, {\cal N}^{3}(r, S_e) \ .
\label{eq:MSA_prob_lethal_final+SW-3}
\end{equation}
This means that the probability of lethal lesions ${\cal P}_{l}(r, S_e) \to 1$ within the entire region where $1 \ll {\cal N}(r, S_e) \ll \nu_{{\rm max}}$.

Knowing ${\cal P}_{l}(r, S_e)$ one can now calculate the number of lethal events in a cell nucleus traversed by a projectile ion. Equation~(\ref{eq:MSA_prob_lethal_final}) represents the probability to create a lethal lesion in a DNA double twist located at the distance $r$ from the ion track.
Integrating ${\cal P}_{l}(r, S_e)$ over the area perpendicular to the ion's trajectory and convoluting the result with the number density of DNA double twists in a cell nucleus one derives the average number of lethal lesions per unit length of the ion's trajectory:
\begin{equation}
\frac{dN_{l}(S_e)}{dx} =
n_s \, \int_0^{\infty} {\cal P}_{l}(r, S_e) \, 2\pi r \, dr  \equiv
n_s \, \sigma_l(S_e) \ .
\label{eq:MSA_eq03}
\end{equation}
Here $n_s$ is the number density of DNA double twists in a cell nucleus which is equal to the number of DNA base pairs
accommodated in a cell nucleus, $N_{{\rm bp}}$, divided by the number of DNA base pairs in one double twist and by the nuclear volume $V_n$ \cite{verkhovtsev2016multiscale},
\begin{equation}
n_s = \frac{ N_{{\rm bp}} }{ 20 \, V_n } \ .
\end{equation}
The function $\sigma_l(S_e)$ is the cross section of producing lethal DNA damage in a cell nucleus, which depends on LET and the concentration of oxygen in the target. The $\sigma_l(S_e)$ dependence originates from the dependence of ${\cal N}(r, S_e)$ on LET; this dependence is discussed further below in this section.

The number of lethal events in a cell nucleus at a given dose $d$ produced by $N_{{\rm ion}}$ ions is equal to \cite{Surdutovich_2014_EPJD.68.353}:
\begin{equation}
Y_{l}(S_e) = \frac{{\rm d}N_{l}(S_e)}{{\rm d}x} \, \bar{z} \, N_{\rm ion}(S_e) \ ,
\label{eq:yield_lethal_nucleus}
\end{equation}
where $\bar{z}$ is the average distance traversed by $N_{{\rm ion}}$ ions through the cell nucleus.
The average number of ions hitting the nucleus, $N_{{\rm ion}}$, depends on the nucleus area $A_n$, the dose and LET:
\begin{equation}
N_{{\rm ion}}(S_e) = A_n \frac{\rho \, d}{S_e} \ ,
\label{eq:number_ions}
\end{equation}
where $\rho$ is the mass density of the irradiated medium taken equal to the density of liquid water, $\rho = 1$~g/cm$^3$.
The probability of cell survival is given by the probability of zero lethal lesions occurrence \cite{Surdutovich_2014_EPJD.68.353}.
According to the Poisson statistics it is equal to
\begin{equation}
\Pi_{\rm surv} = e^{-Y_l(S_e)} \ .
\label{eq:surv_prob}
\end{equation}
Substituting $N_{{\rm ion}}$ into $Y_{l}$ and taking the logarithm of $\Pi_{\rm surv}$ one obtains
\begin{equation}
\ln {\Pi_{{\rm surv}} } = -Y_l(S_e) = - \frac{{\rm d}N_{l}(S_e)}{{\rm d}x} \, \bar{z} \, A_n \frac{\rho \, d}{S_e} \ ,
\label{eq_cell_surv_prob}
\end{equation}
where $\left( \frac{{\rm d}N_{l}}{{\rm d}x} \, \bar{z} \right)$ is the average number of lethal events created by a single ion in a cell nucleus.

Now let us analyze the dependence of the cross section of a DNA lethal lesion $\sigma_l$ and the number of lethal lesions in a cell nucleus $Y_{l}$ on LET.
First, consider the irradiation with low-LET ions at the Bragg peak region. A representative case for this scenario is irradiation with protons. In this case the number of lesions in a DNA double twist ${\cal N}(r, S_e) \ll 1$ and hence ${\cal P}_{l}(r, S_e) \sim {\cal N}^{3}(r, S_e)$ according to Eq.~(\ref{eq:MSA_prob_lethal_final+SW-2a}).
For protons at the Bragg peak, the number of lesions in a DNA double twist is proportional to LET, ${\cal N}(r, S_e) \propto S_e$. This dependence can be explained as follows.
The number of lesions created by secondary electrons incident on a DNA double twist, ${\cal N}_{\rm e}(r, S_e)$, is proportional to the number of such electrons (see Eq.~(\ref{eq:MSA_numlesions_Ne})), which in turn is proportional to LET.
The number of lesions created by free radicals, ${\cal N}_{\rm r}(r, S_e)$, is determined by the number of such species, which is proportional to the number of secondary electrons \cite{surdutovich2015transport}.
The shock wave induced by protons at the Bragg peak does not transport free radicals and other reactive species to the distances much larger than the secondary electron propagation range $R_{{\rm e}}$. Note, however, that the diffusion of free radicals on the picosecond time scale might be affected by the temperature increase in the vicinity of the ion tracks.
As follows from the analysis described below in Sect.~\ref{sec:SW_propagation}, the free radicals propagation range $R_{{\rm r}}$ is smaller than $R_{{\rm e}} \sim 1-2$~nm in the $S_e$ region up to 70--140~keV/$\mu$m.
Combining Eqs.~(\ref{eq:MSA_prob_lethal_final+SW-2a}) and (\ref{eq:MSA_eq03}) and using the ${\cal N}(r, S_e) \propto S_e$ dependence one obtains that in this case $\sigma_l$ depends on LET as
\begin{equation}
\sigma_l(S_e) \propto S_e^3 \ .
\label{eq:MSA_sigma_l-protons}
\end{equation}
The number of lethal lesions in a cell nucleus $Y_{l}$, Eq.~(\ref{eq:yield_lethal_nucleus}), thus increases with LET as
\begin{equation}
Y_{l}(S_e) \propto \frac{\sigma_l(S_e)}{S_e} \sim S_e^{2} \ .
\label{eq:MSA_sigma_2_protons}
\end{equation}

The quantity ${\cal N}_{\rm r}(r, S_e)$ might grow with the growth of LET due to the increase of the SW radius and correspondingly $R_{{\rm r}}$. The growth of $R_{{\rm r}}$ results in lowering the density of free radicals and thus their recombination rate constant. The additional growth of ${\cal N}_{\rm r}(r, S_e)$ with $S_e$ will result in the faster growth of $\sigma_l$ and $Y_l$ with increasing $S_e$.
Even steeper dependencies of $\sigma_l$ and $Y_l$ on LET may arise at higher $S_e$ values when the number of lesions in a DNA double twist ${\cal N}(r, S_e) \lesssim 1$ due to a steeper dependence of ${\cal P}_{l}(r, S_e)$ on ${\cal N}(r, S_e)$, see Fig.~\ref{fig:prob_lethal_calN}.

Finally, let us consider the case ${\cal N}(r, S_e) \gg 1$ when the probability ${\cal P}_{l}(r, S_e) \to 1$.
This case describes iron and heavier ions at the Bragg peak. In this case multiple lesions are created by the shock wave induced thermomechanical stress of the DNA double twist within the distance range  $r <  R_{{\rm SW}}(S_e)$ from the ion track.
The number of lesions produced by the shock wave in the region $r <  R_{{\rm SW}}(S_e)$ is much bigger than the number of lesions produced by secondary electrons and free radicals, i.e. ${\cal N}_{{\rm SW}}(r, S_e) \gg {\cal N}_{{\rm e}}(r, S_e)$ and ${\cal N}_{{\rm SW}}(r, S_e) \gg {\cal N}_{{\rm r}}(r, S_e)$.
As described in detail in Sect.~\ref{sec:Number_bond_breaks_MD}, the number of lesions ${\cal N}_{{\rm SW}}(r, S_e)$ has been evaluated from the MD simulations for the five ions with different LET values at the Bragg peak region. The critical distance $R_{{\rm SW}}$ is analyzed below in Sect.~\ref{sec:rupture_force}.
These results suggest the following stepwise dependence of ${\cal N}(r, S_e)$ on distance $r$ from the ion track:
\begin{equation}
{\cal N}(r, S_e) = {\cal N}_{{\rm SW}}(S_e) \, \theta( R_{{\rm SW}}(S_e) - r ) \ .
\label{eq:MSA_numlesions_N_SW}
\end{equation}

Since at large LET values ${\cal N} \approx {\cal N}_{{\rm SW}} \gg 1$ within the range $r <R_{{\rm SW}}(S_e)$, the probability ${\cal P}_{l}(r, S_e) \to 1$ at $r < R_{{\rm SW}}(S_e)$. Then for high-LET irradiation one obtains
\begin{equation}
\sigma_l(S_e) =  \int_0^{\infty} {\cal P}_{l}(r, S_e) \, 2\pi r \, dr  = \pi R_{{\rm SW}}^2(S_e) \ .
\end{equation}
In this case the number of lethal events in a cell nucleus at a given dose $d$ produced by $N_{{\rm ion}}$ ions, Eq.~(\ref{eq:yield_lethal_nucleus}), transforms into:
\begin{equation}
Y_l(S_e) = \pi R_{{\rm SW}}^2(S_e) \, n_s \, \bar{z} \,  A_n \frac{\rho \, d}{S_e}
\label{eq:yield_lethal_nucleus+SW}
\end{equation}
with the probability of cell survival being given by Eq.~(\ref{eq:surv_prob}).
As demonstrated below in Section~\ref{sec:rupture_force}, $R_{{\rm SW}}$ depends on LET as $R_{{\rm SW}} = b \, S_e^{1/3}$, where $b$ is the proportionality factor determined in Sect.~\ref{sec:rupture_force}. In the case of large LET values (where the condition ${\cal N} \approx {\cal N}_{{\rm SW}} \gg 1$ is fulfilled) the number of lethal events in a cell nucleus can be written as
\begin{equation}
Y_l(S_e) = \alpha \, S_e^{-1/3}
\label{eq:MSA_yield_highLET}
\end{equation}
where
\begin{equation}
\alpha = \pi \, b^2 \, n_s \, \bar{z} \, A_n \rho \, d \ .
\end{equation}
This means that the number of lethal events in a cell nucleus at a given dose $d$ decreases slowly with high LET, which corresponds to
the experimental observations for iron and heavier ions at the Bragg peak region, see Sect.~\ref{sec:Results_cell-surv}.

\section{Results and discussion}
\label{sec:Results}

The first part of this section presents the results of the reactive MD simulations of the shock wave induced damage occurring in a 30 base pairs long DNA segment introduced in Fig.~\ref{fig. parameters}. The simulations revealed that most bond breaks in the DNA backbone are produced within the central segment consisting of two helical turns and containing 20 base pairs. Therefore, the number of bond breaks in the central DNA double twist, being the target DNA segment considered within the MSA formalism \cite{Surdutovich_2014_EPJD.68.353}, has been quantified.
The shock wave induced dynamics of the liquid water medium is analyzed next to evaluate the range of shock wave propagation and hence the range of shock wave driven propagation of reactive species.
The analysis concludes with the evaluation of survival probabilities of cells irradiated with high-LET ions within the MSA formalism. This analysis reveals the significant role of the shock wave induced thermomechanical mechanism of DNA damage in the cell inactivation.

\subsection{Quantification of the number of bond breaks in the DNA double twist}
\label{sec:Number_bond_breaks_MD}

\begin{figure}[t!]
\centering
\includegraphics[width=0.95\textwidth]{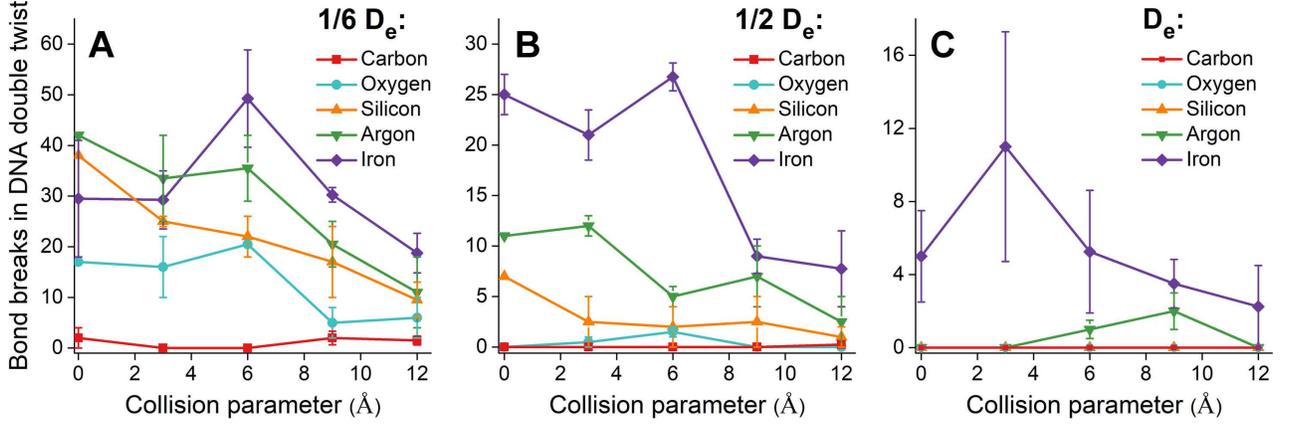}
\caption{Average number of bond breaks in the DNA double twist calculated as a function of the collision parameter $d_{\text{geo}}$ for the five studied ions. $d_{\text{geo}}$ is the distance from the ion track to the principal axis of inertia of the DNA segment as shown in Fig.~\ref{fig. parameters}A. Panels A, B and C show the results of simulations employing the bond dissociation energies $D_e$ derived from the DFT calculations (see Fig.~\ref{fig. parameters}B) and the values of $D_e$ scaled by the factors of 1/2 and 1/6. Two independent MD simulations have been performed for each projectile ion and each collision geometry; error bars indicate the corresponding standard deviation.}
\label{fig:probability of ssb}
\end{figure}

MD simulations of the shock wave induced damage of a 30 base pairs long DNA molecule reveal that the projectile ion propagating in close proximity to the principal axis of inertia of the molecule produces significant damage within the central segment containing 20 DNA base pairs.
The DNA damage produced in segments of such size may lead to complex irreparable lesions in a cell \cite{Surdutovich_2014_EPJD.68.353, verkhovtsev2016multiscale,schipler2013dna}.
The number of bond breaks in the DNA double twist was counted after each completed MD simulation and analyzed as a function of the distance $d_{\text{geo}}$ from the ion's path to the principal axis of inertia of the DNA molecule, see Fig.~\ref{fig. parameters}A.
The results of the performed analysis are shown in Fig.~\ref{fig:probability of ssb}.

The figure shows that the number of bond breaks produced in the DNA double twist by the ion-induced shock wave increases with the LET of a projectile ion (see Table~\ref{tab: LET_BP}). Simulation results obtained for the scaled bond dissociation energies $D_e/6$ (Fig.~\ref{fig:probability of ssb}A) reveal that up to two DNA backbone bonds break due to the shock wave induced by the carbon ion whereas up to 50 bonds may be broken due to the iron ion impact.
For every combination of the bond dissociation energy and ion's LET the average number of bond breaks fluctuates around certain values ${\cal N}_{{\rm SW}}$ within a certain distance range from the ion's path; the values ${\cal N}_{{\rm SW}}$ for different LET values are summarized in Table~\ref{tab:Numbreaks_LET_De}.
As the ion passes at larger distances from the main axis of the DNA molecule the average number of bond breaks within the DNA double twist gradually decreases.
Figure~\ref{fig:probability of ssb} shows that the shock wave induced thermomechanical stress of the DNA mostly occurs at $\lesssim 1$~nm from the ion's path for ions lighter than iron.
A more systematic and precise analysis of the threshold distance from the ion's path for inducing DNA strand breaks for each projectile ion is possible, but it would require a significantly larger number of additional simulations aiming at decreasing statistical uncertainties and considering larger values of the collision parameter $d_{\text{geo}}$. Such an analysis might be considered in the future.

\begin{table}[t!]
\centering
\caption{Characteristic number of bond breaks, ${\cal N}_{{\rm SW}}$, occurring in the DNA double twist within a certain distance range $r_0$ from the ion's path to the principal axis of inertia of the DNA segment. Different columns correspond to the results of simulations where the default bond dissociation energies $D_e$ \cite{friis_2020} as well as the scaled bond dissociation energies $D_e/2$ and $D_e/6$ were used, as described in Sect.~\ref{sec:simulation_parameters}.}
\begin{tabular}{p{1.8cm} | p{1.9cm}p{1.5cm}p{0.01cm} | p{1.9cm}p{1.5cm}p{0.01cm} | p{1.9cm}p{1.5cm}}
\hline
         & \multicolumn{2}{c}{$D_e/6$}  & & \multicolumn{2}{c}{$D_e/2$} & & \multicolumn{2}{c}{$D_e$} \\
\hline
   &  ${\cal N}_{{\rm SW}}$   &  $r_0$~(nm) & &   ${\cal N}_{{\rm SW}}$   & $r_0$~(nm) & &  ${\cal N}_{{\rm SW}}$ & $r_0$~(nm)  \\
\hline
Carbon   &      $1.1 \pm 0.5$   &     1.2      & &     $0.05 \pm 0.05$ & 1.2        & &      0            & 1.2   \\
Oxygen   &      $17.8 \pm 1.4$  &     0.6      & &     $0.7 \pm 0.4$   & 0.6        & &     0            & 1.2   \\
Silicon  &      $28.3 \pm 4.9$  &     0.9      & &     $3.5 \pm 1.2$   & 0.9        & &     0            & 1.2   \\
Argon    &      $32.9 \pm 4.5$  &     0.9      & &     $8.8 \pm 1.7$   & 0.9        & &    $0.8 \pm 0.5$ & 0.9   \\
Iron     &      $34.0 \pm 4.9$  &     0.9      & &     $24.2 \pm 1.7$  & 0.9        & &    $5.4 \pm 1.5$ & 1.2   \\
\hline
\end{tabular}
\label{tab:Numbreaks_LET_De}
\end{table}

\begin{figure}[!tb]
\centering
\includegraphics[trim=0cm 2cm 0cm 0.5cm,width=0.9\textwidth]{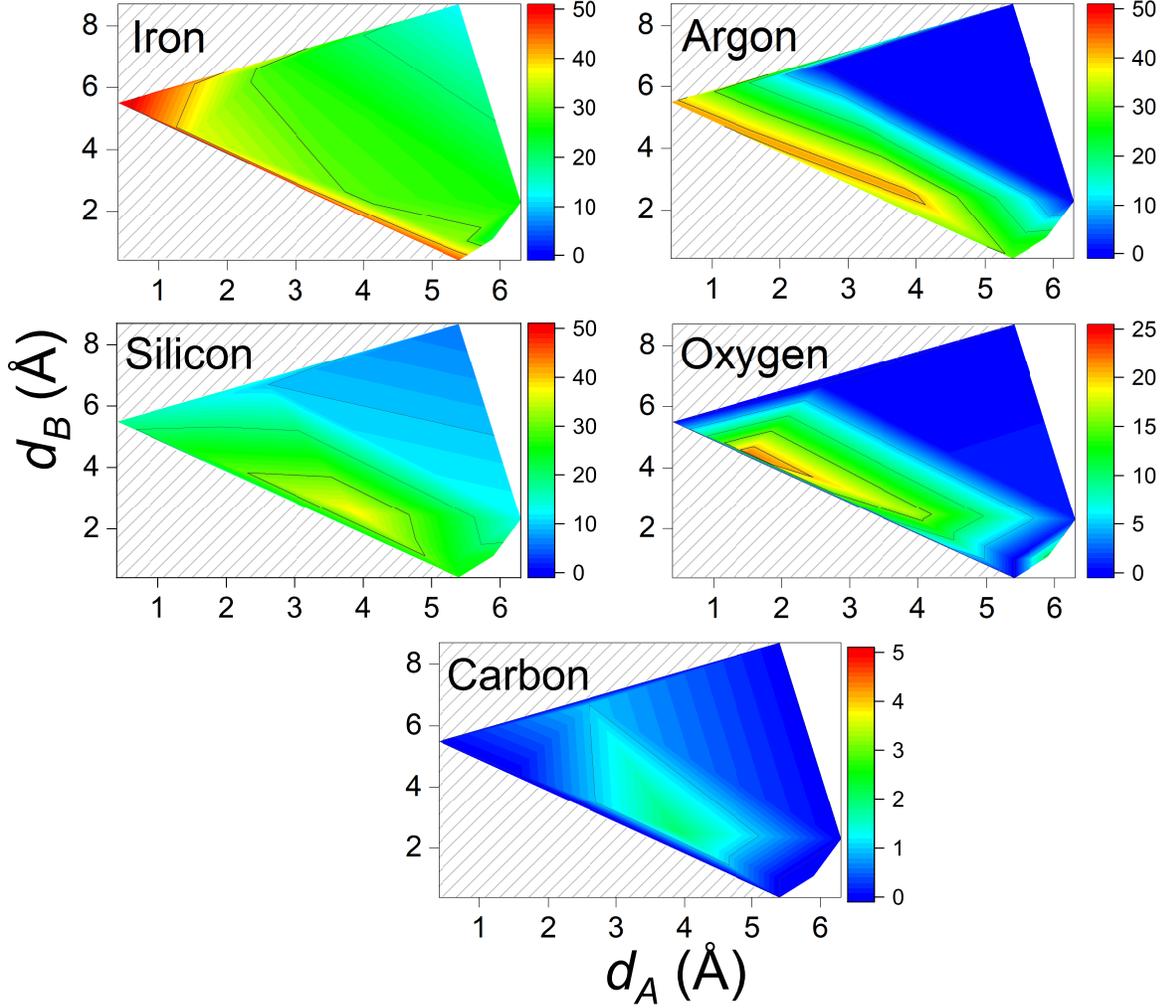}
\caption{Spatial distribution of the total number of bond breaks occurring in a DNA double twist (color gradient) as a function of two collision parameters $d_A$ and $d_B$ (see Fig.~\ref{fig. parameters}), computed for Fe, Ar, Si, O and C ions in the Bragg peak region. The spatial region inaccessible for the given combination of collision parameters is marked with dashed grey lines. Note that the scale of the color bars for Fe, Ar and Si ions is twofold larger than for the O ion and tenfold larger than for the C ion.}
\label{fig:probability of dsb}
\end{figure}

The spatial distribution of the total number of bond breaks occurring in the DNA double twist has been analyzed as a function of the shortest distances $d_A$ and $d_B$ (see Fig.~\ref{fig. parameters}) from the ion's path to DNA strand A and strand B, respectively. Figure~\ref{fig:probability of dsb} shows the results obtained with the scaled bond dissociation energies $D_e/6$.
Due to the geometry of the studied system, not all combinations of $d_A$ and $d_B$ are accessible at the same time; the inaccessible spatial regions in Fig.~\ref{fig:probability of dsb} are marked with dashed lines.
The total number of strand breaks in the DNA double twist decreases with the simultaneous increase of the distance from the ion track to both DNA strands. The results shown in Fig.~\ref{fig:probability of dsb} indicate that the number of bond breaks drops sharply when the ion track passes at distances larger than 5~\AA~to both DNA strands. In contrast, the largest number of bond breaks in the sugar-phosphate backbone is observed when the ion passes in close proximity to at least one of the DNA strands.

\subsection{Propagation range for ion-induced shock waves}
\label{sec:SW_propagation}

\begin{figure}[t!]
\centering
\includegraphics[width=0.5\textwidth]{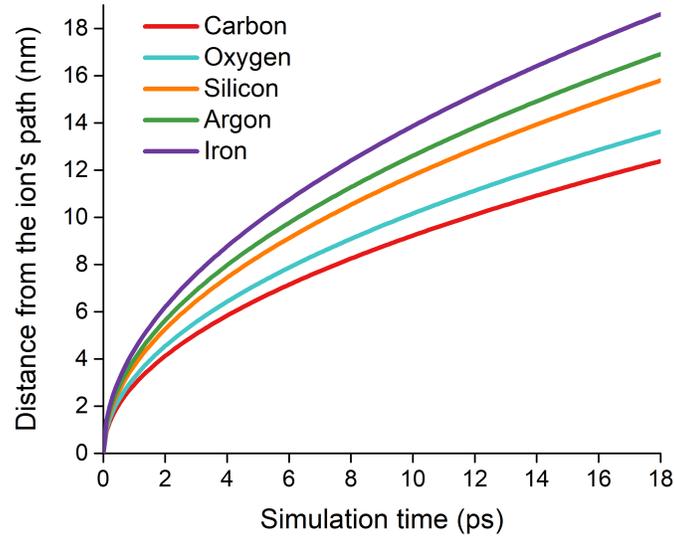}
\caption{Radial distance travelled by the front of the shock wave induced by the five studied ions in the Bragg peak region. The results are obtained using Eq.~(\ref{eq. shock wave propagation}).}
\label{fig:density wave}
\end{figure}

The front of the shock wave propagates radially away from the ion's path.
The dependence of the radius of the wave front $R$ on time reads as \cite{surdutovich2010shock}:
\begin{equation}
R = \beta \sqrt{t} \left( \frac{S_e}{\rho} \right)^{1/4} \ ,
\label{eq. shock wave propagation}
\end{equation}
where $t$ is the time from the start of the shock wave propagation, $S_e$ is the ion's LET, $\rho$ is the density of the unperturbed medium ($\rho = 1$~g/cm$^3$ for liquid water) and $\beta = 0.86$ is a dimensionless parameter determined in \cite{surdutovich2010shock}.
The position of the wave front calculated for different projectile ions is illustrated in Fig.~\ref{fig:density wave}.

\begin{figure}[t!]
\centering
\includegraphics[width=\textwidth]{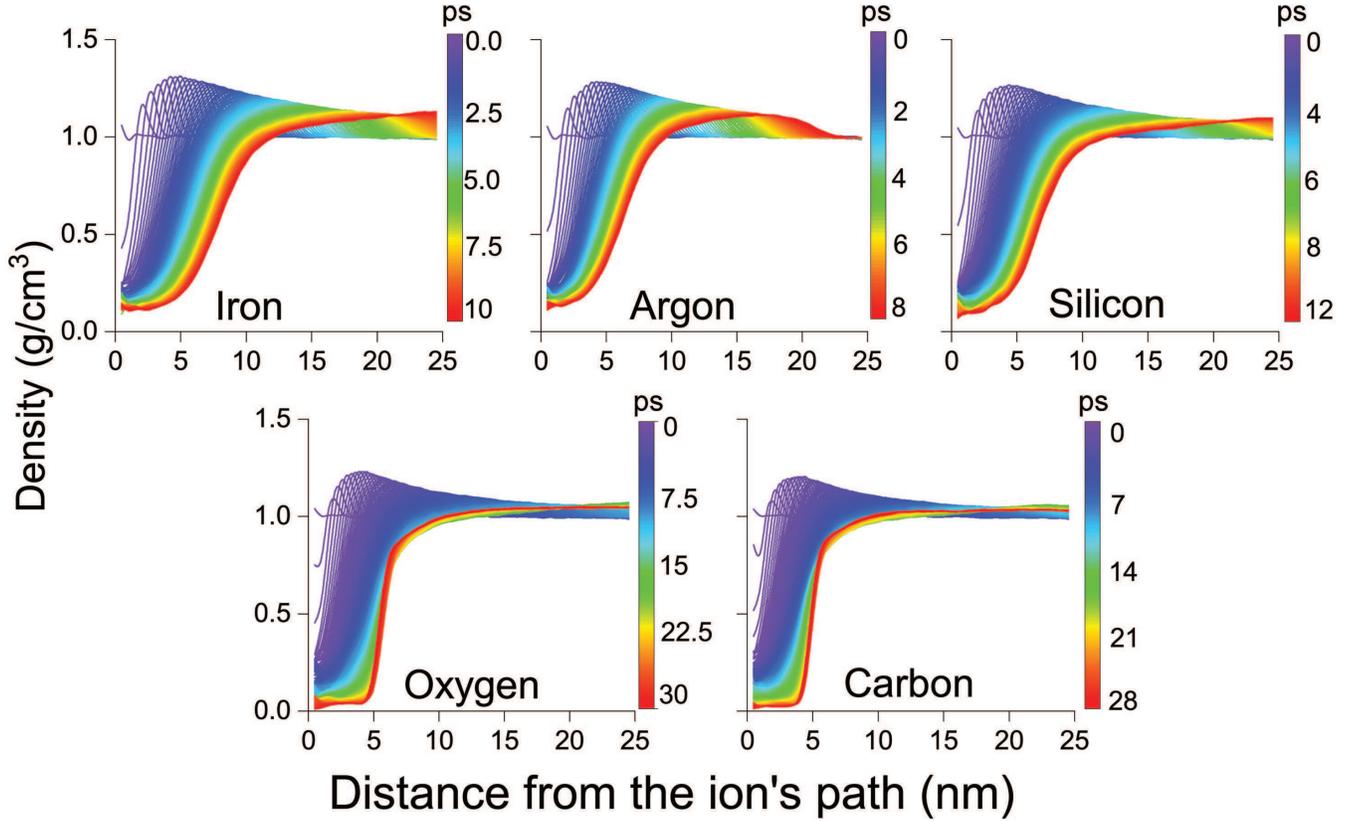}
\caption{The average density of water as a function of radial distance from the ion track. The dynamics of the water medium is caused by propagation of the shock wave induced by iron, argon, silicon, oxygen and carbon ions inside a $49.5~\textrm{nm} \times 49.5~\textrm{nm} \times 8.0~\textrm{nm}$ water box. The simulation time (measured in ps) is depicted as a color scale. The shock wave induced by iron, argon and silicon ions (top row) has reached the simulation box boundary much faster than the shock wave induced by oxygen and carbon ions (bottom row). Therefore, the simulation time for iron, argon and silicon ions is about 3 times shorter than for the lighter ions. }
\label{fig:water shock wave}
\end{figure}

To evaluate the range of shock wave driven propagation of reactive species, a shock wave propagation was simulated in a pure water box with dimensions of $49.5~\textrm{nm} \times 49.5~\textrm{nm} \times 8.0~\textrm{nm}$.
The evolution of radial density of water around the tracks of the five projectile ions is shown in Fig.~\ref{fig:water shock wave}.
Water molecules located in the vicinity of the ion's path are transported away from their initial positions, which results in the formation of a cylindrical cavity around the ion's path. The radius of the cavity grows with time up to the values of about 6~nm for carbon and oxygen ions, while the density of water increases at larger distances from the ion's path.
Following the mass conservation law, the mass of water molecules transported from the region in the vicinity of the ion track should be equal to the mass of excess water molecules at larger distances from the track.

\begin{figure}[t!]
\centering
\includegraphics[width=0.9\textwidth]{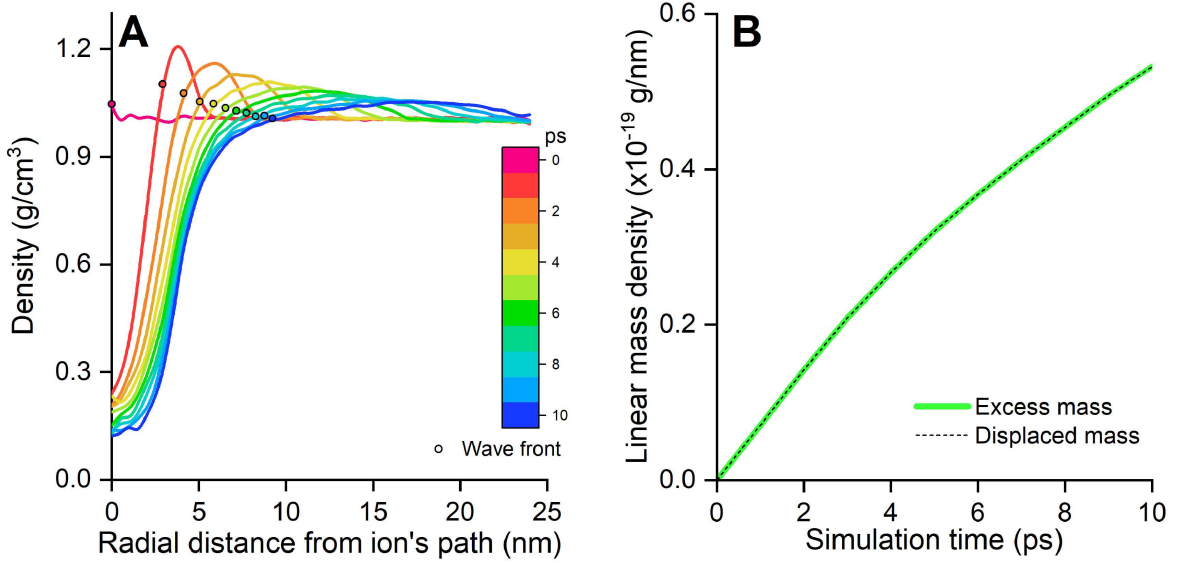}
\caption{\textbf{A:} Variation of the density of water caused the carbon ion induced shock wave as a function of radial distance from the ion track. The simulation time is indicated in the sidebar. The position of the shock wave front $R$ calculated using Eq.~(\ref{eq. shock wave propagation}) is indicated with dots. The position of the wave front separates the mass transported by the shock wave, that is considered as displaced (behind the wave front, i.e. at $r < R$) and excess (beyond the wave front, i.e. at $r > R$).
\textbf{B:} Time evolution of the linear mass density, defined as the water mass transported by the carbon ion-induced shock wave and normalized by the unit length of ion's trajectory. The displaced mass calculated behind the wave front is shown by the dashed black line whereas the excess mass beyond the wave front is shown by the solid green line. }
\label{fig: mass conservation}
\end{figure}

The position of the shock wave front at different time instances, determined using Eq.~(\ref{eq. shock wave propagation}), is depicted in Fig.~\ref{fig: mass conservation}A with symbols for the case of a carbon ion induced shock wave. The mass of all water molecules behind the wave front is calculated and normalized to the unit length of ion's trajectory. The obtained value is compared to the normalized excess mass beyond the wave front. The comparison shown in Fig.~\ref{fig: mass conservation}B illustrates that the linear mass density is indeed conserved within the simulation time range considered.

\begin{figure}[!tb]
\centering
\includegraphics[width=0.9\textwidth]{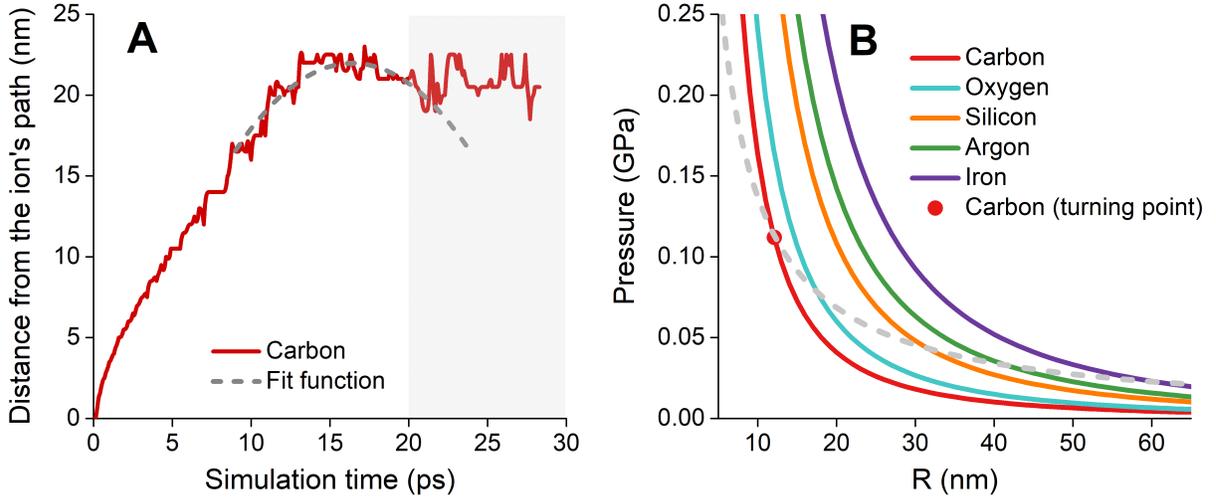}
\caption{\textbf{A:} The radial position of the maximal density of water as a function of simulation time for the shock wave induced by a carbon ion in the Bragg peak region. The maximal distance of the shock wave propagation is defined by a quadratic fit function (see the dashed line). The nonphysical region due to the shock wave reflection from the simulation box boundaries is marked with grey color.
\textbf{B:} The pressure exerted by the shock wave front generated by different ions in the Bragg peak region as a function of the wave front radius $R$. The red dot depicts the maximal propagation distance for the shock wave front generated by a carbon ion (the ``turning point''), calculated using Eq. (\ref{eq. shock wave propagation}). The corresponding time instance, $t = 16.7$~ps, has been determined from the MD simulations as shown in panel~A. The dashed line shows the surface tension pressure on the surface of a cylindrical wake region with radius $R$.}
\label{fig:pressure_turning_point}
\end{figure}

The simulation of the propagation of the carbon ion induced shock wave reveals that at a certain time instance the shock wave has stopped propagating away from the ion's path and started to move slowly in the inward direction. This happens when the pressure of the shock wave front drops below a certain value determined by the balance of the pressure at the wave front and the water surface tension pressure \cite{surdutovich2017ion, ES_AVS_2019_EPJD.73.241}.
Fig.~\ref{fig:pressure_turning_point}A shows the radial position of the maximal density of water as a function of simulation time for the case of the projectile carbon ion. The radial displacement of the maximal density from the ion track axis increases rapidly during the first 15~ps of the simulation, then reaches the maximal value and starts to decrease at later time instances. The analysis shown in Figure~\ref{fig:pressure_turning_point}A suggests that the maximal radial displacement of the density corresponds to the time instance $t = 16.7$~ps. Note that for $t > 20$~ps the radial displacement of the maximal density stops decreasing but fluctuates around the value of 21~nm. This behavior is attributed to interference with the outer part of the shock wave front, which reaches the simulation box boundary and gets reflected. The behavior of the system within the simulation time range $t \le 20$~ps is nevertheless physically meaningful as the shock wave has not yet reached the simulation box boundary within this time interval.

According to Eq.~(\ref{eq. shock wave propagation}), the front of the carbon ion induced shock wave propagates by the time $t = 16.7$~ps to the distance $R = 11.9$~nm from the ion track. This characteristic distance defines the propagation range of free radicals, $R_{{\rm r}}$ in Eq.~(\ref{eq:MSA_numlesions_Nr}), which are transported by the shock wave driven collective flow.
To evaluate the range of shock wave propagation for heavier ions one would need to run longer simulations and consider much larger simulation boxes than the one used in the present study.
Alternatively, the range of the shock wave propagation induced by high-LET ions can be estimated from the analysis of the pressure on the shock wave front \cite{surdutovich2010shock}:
\begin{equation}
P = \frac{ \beta^4 }{2(\gamma + 1)} \frac{S_e}{R^2} \ ,
\label{eq:MSA_SW_front_pressure}
\end{equation}
where $\beta = 0.86$ is a dimensionless parameter determined in \cite{surdutovich2010shock}, $\gamma = C_P/C_V$ is the heat capacity ratio ($\gamma = 1.222$ for liquid water), and $R$ is the radius of the shock wave front. Figure~\ref{fig:pressure_turning_point}B shows the pressure induced by the shock wave front generated by the different ions in the Bragg peak region. Colored lines correspond to the results derived using Eq.~(\ref{eq:MSA_SW_front_pressure}). A red dot depicts the pressure $P = 0.115$~GPa at the distance $R = 11.9$~nm from the ion track, that is the maximal distance of the wave front propagation for a carbon ion. The indicated value of $R$ corresponding to the instant $t = 16.7$~ps has been evaluated using Eq.~(\ref{eq. shock wave propagation}).

\begin{table}[t!]
\centering
\caption{The maximum radii of the shock wave wake region for the five studied ions at their Bragg peak energies. These characteristic radii define the propagation range of free radicals, $R_{{\rm r}}$, which are transported by the propagating shock wave.}
\begin{tabular}{p{2.0cm}p{1.5cm}p{1.5cm}p{1.5cm}p{1.5cm}p{1.5cm}}
\hline
                   &  Carbon  &  Oxygen  &  Silicon  &  Argon  &  Iron  \\
\hline
 $R_{{\rm r}}$~(nm) &   11.9   &   17.5   &   31.6    &   41.5  &  60.8   \\
\hline
\end{tabular}
\label{tab:Critical_Pressure}
\end{table}

The shock wave propagation in the radial direction away from the ion's path causes cavitation in its wake, leading to the formation of a rarefied cylindrical region \cite{surdutovich2010shock, surdutovich2017ion}. This effect has been observed in MD simulations described in Fig.~\ref{fig:water shock wave} and Fig.~\ref{fig: mass conservation}A.
In the course of the shock wave propagation, the pressure at the shock wave front becomes balanced by the surface tension pressure building up at the border of the wake region. As a result, the growth of the wake region stops and this region shrinks after the instant when the pressure of the wave front becomes equal to the water surface tension pressure on the surface of the wake region. The latter can be estimated as
\begin{equation}
P = \frac{\xi}{R} \ ,
\label{eq:pressure_surf_tension}
\end{equation}
where $\xi$ is the coefficient of surface tension and $R$ is the distance from the ion track. Using the aforementioned values $P = 0.115$~GPa and $R = 11.9$~nm for the carbon ion at the Bragg peak, one obtains the surface tension coefficient $\xi = 1.37$~N/m.
Note that the medium in the vicinity of the shock wave front is far from equilibrium, and the density of the medium is significantly higher than the density of water at ambient conditions.
The high water density in the vicinity of the shock wave front and the large amount of energy deposited into the medium explain the large value of the corresponding surface tension coefficient.
The dependence of the surface tension pressure on the distance from the ion track, calculated using Eq.~(\ref{eq:pressure_surf_tension}), is shown in Fig.~\ref{fig:pressure_turning_point}B by a dashed line.

Assuming that $\xi$ depends weakly and smoothly on LET (or does not depend at all) at the pressures balance point one can evaluate the radii $R_{{\rm r}}$ for different ions. The radii $R_{{\rm r}}$ define the propagation ranges of free radicals transported by the shock wave induced by different ions.
Equating the pressure on the shock wave front, Eq.~(\ref{eq:MSA_SW_front_pressure}), and the surface tension pressure, Eq. (\ref{eq:pressure_surf_tension}), one obtains a linear dependence of $R_{{\rm r}}$ on LET:
\begin{equation}
R_{{\rm r}} = \frac{ \beta^4 }{2(\gamma + 1) \, \xi} \, S_e \ .
\label{eq:R_r_linear_LET}
\end{equation}
The calculated values of $R_{{\rm r}}$ for the five studied ions at the Bragg peak region are summarized in Table~\ref{tab:Critical_Pressure}.
The results indicate that for an iron ion the free radicals are transported by the shock wave to the distance of $\sim$60~nm.
This value exceeds by an order of magnitude typical distances that radicals can diffuse in the medium being at the equilibrium during the time corresponding to the duration of formation of the shock wave wake region with the maximum radius.

\subsection{Force exerted by the shock wave on the DNA}
\label{sec:rupture_force}

The characteristic range of the shock wave induced thermomechanical damage, $R_{{\rm SW}}$, can be evaluated by analyzing the pressure on the shock wave front, Eq.~(\ref{eq:MSA_SW_front_pressure}), and the corresponding force exerted by the shock wave on covalent bonds in the DNA backbone.

Equation~(\ref{eq:MSA_SW_front_pressure}) arises as a solution of the hydrodynamic Euler equation, the continuity and entropy conservation equations \cite{surdutovich2010shock}.
As such, it is applicable to the volumes with sizes being much larger than the characteristic size of a single molecule or a molecular bond. Nevertheless, the medium pressure induces forces applied to single molecular bonds. Such forces should be treated as a result of averaging over an ensemble of molecules exposed to the pressure $P(r)$ at a distance $r$ from the ion track.

Let us evaluate forces acting on molecular bonds in the presence of the ion-induced shock wave. For the sake of simplicity, let us consider a molecular bond oriented parallel to the direction of the shock wave propagation. The force stretching the bond is given by:
\begin{equation}
F = (\pi a_0^2) \, \frac{\partial P}{\partial r} \, l  \ ,
\label{eq:stretching_force}
\end{equation}
where $\pi a_0^2$ is the transverse area of the molecular bond exposed to the pressure created by the shock wave front,
$\frac{\partial P}{\partial r}$ is the pressure gradient, and
$l$ is a characteristic interatomic distance in the medium on which the pressure gradient is evaluated.
The calculations described below are performed using the value $a_0 = 0.15$~nm corresponding to the van der Waals radius for atoms forming the DNA backbone \cite{Bondi_1964_vdW_radii_JPC.68.441} and $l \approx 0.15$~nm, that is a characteristic length of covalent bonds in the DNA backbone.

Let us consider potential energy of a DNA backbone bond being under the pressure created by the shock wave front:
\begin{equation}
U(r) = D_e \left[ e^{-2\kappa (r - r_0)} - 2 e^{-\kappa (r - r_0)} \right] - \int_{r_0}^{r} \vec{F} \cdot {\rm d}\vec{r} \ .
\label{eq:bond_rupture-1new}
\end{equation}
The first term on the right-hand side of Eq.~(\ref{eq:bond_rupture-1new}) is the bond potential energy described by the Morse potential, $D_e$ is the bond dissociation energy, $r_0$ is the equilibrium bond length, and $\kappa$ defines the steepness of the potential energy curve.
These parameters for the C$_3^{\prime}$--O, C$_4^{\prime}$--C$_5^{\prime}$, C$_5^{\prime}$--O and P--O bonds in the DNA backbone (see Fig.~\ref{fig. parameters}B) are determined from the potential energy curves obtained by means of DFT \cite{friis_2020} and listed in Table~\ref{tab:backbone_bonds_param}.
The second term on the right-hand side of Eq.~(\ref{eq:bond_rupture-1new}) describes the work against the force $\vec{F}$, caused by the pressure gradient on the distance from $r_0$ to $r$.
The force $\vec{F}$ should not depend on the interatomic distance $r$ as it originates from the medium and is defined by its properties at given thermodynamic conditions and at any given location of the bond in space.
Considering the geometry when the bond is oriented along $\vec{F}$ one derives
\begin{equation}
U(r) = D_e \left[ e^{-2\kappa (r - r_0)} - 2 e^{-\kappa (r - r_0)} \right] - F \times (r - r_0) \ .
\label{eq:bond_rupture-1}
\end{equation}

\begin{table}[t!]
\centering
\caption{The bond dissociation energy $D_e$, the equilibrium bond length $r_0$ and steepness of the potential energy curve $\kappa$ for the C$_3^{\prime}$--O, C$_4^{\prime}$--C$_5^{\prime}$, C$_5^{\prime}$--O and P--O bonds in the DNA backbone (see Fig.~\ref{fig. parameters}B). These parameters have been determined from the potential energy curves obtained by means of DFT \cite{friis_2020}.}
\begin{tabular}{p{2.0cm}p{1.5cm}p{1.5cm}p{1.5cm}p{1.5cm}}
\hline
                     &  C$_3^{\prime}$--O  &  C$_4^{\prime}$--C$_5^{\prime}$  &  C$_5^{\prime}$--O  &   P--O  \\
\hline
$D_e$~(eV)           &        6.95         &               6.36               &          5.90       &   6.33   \\
$r_0$~(\AA)          &        1.42         &               1.52               &          1.45       &   1.61   \\
$\kappa$~(nm$^{-1}$) &       14.54         &              13.11               &         15.00       &  13.91   \\
\hline
\end{tabular}
\label{tab:backbone_bonds_param}
\end{table}

\begin{figure}[t!]
\centering
\includegraphics[width=0.48\textwidth]{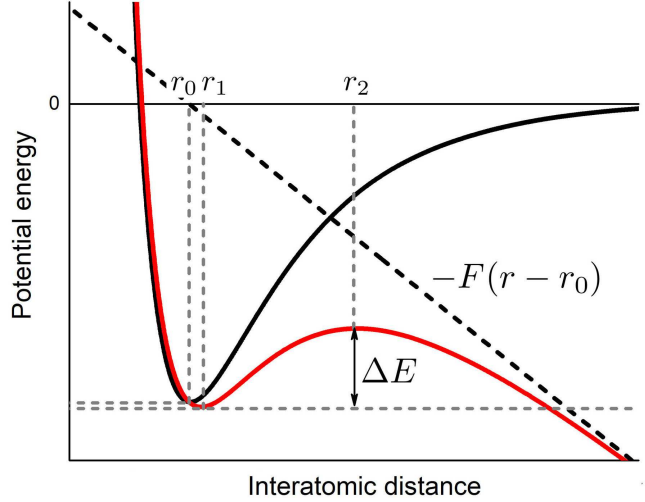}
\caption{Variation of the potential energy curve for a molecular bond upon the action of the external force $F$ caused by the pressure gradient due to the ion-induced shock wave. Solid black line shows the bond potential energy described by the Morse potential. The potential $U(r)$, Eq.~(\ref{eq:bond_rupture-1}), is shown by solid red line. $\Delta E$ is the energy barrier for bond rupture by the shock wave induced thermomechanical stress. See the main text for details.}
\label{fig:SW_forces_schematic}
\end{figure}

Stretching the DNA backbone bond by the force $F$ results in lowering the energy barrier for bond rupture, see Fig.~\ref{fig:SW_forces_schematic}. The energy barrier height reads as
\begin{equation}
\Delta E = U(r_0 + \Delta r_2) - U(r_0 + \Delta r_1)  \ ,
\label{eq:bond_rupture-2}
\end{equation}
where $\Delta r_1 = r_1 - r_0$ is the shift of the potential energy minimum with respect to $r_0$, and $r_2 = r_0 + \Delta r_2$ is the position of the potential energy maximum, see Fig.~\ref{fig:SW_forces_schematic}.

The threshold value of the external force at which the bond rupture becomes possible depends on the amount of energy accessible for atoms forming the bond at a given temperature $T$ and on the amount of energy deposited into the medium by the projectile ion.
The condition for the bond rupture due to the shock wave induced thermomechanical stress of the DNA can be formulated as follows:
\begin{equation}
2 \, \frac{k_{\rm B} T}{2} + \frac{\mu \left( \Delta \vec{v} \right)^2}{2} \ge \Delta E \ .
\label{eq:bond_rupture-6}
\end{equation}
The first term on the left-hand side of Eq.~(\ref{eq:bond_rupture-6}) is the average energy available for one degree of freedom in a thermodynamic system being at the equilibrium at $T = 300$~K.
The factor 2 arises since both kinetic and potential energies of the bond are equal to $\frac12 k_{\rm B}T$ according to the equipartition theorem. The second term on the left-hand side is the kinetic energy of the relative interatomic motion caused by the shock wave; $\mu = m_i m_j/(m_i + m_j)$ is the reduced mass of a pair of atoms $i$ and $j$. The analysis described below is performed for the C$_3^{\prime}$--O, C$_4^{\prime}$--C$_5^{\prime}$, C$_5^{\prime}$--O and P--O bonds in the DNA backbone which are shown in Fig.~\ref{fig. parameters}B.

The gradient of the pressure created by the shock wave on the radial distance $l$ results in the variation of the relative velocity between the atoms forming a molecular bond in the DNA backbone. This variation can be calculated as follows:
\begin{equation}
\Delta \vec{v} =
\vec{v}_j(\vec{r}+\vec{l}) - \vec{v}_i(\vec{r}) =
\vec{v}(\vec{r}+\vec{l}) + \vec{v}_j^{\prime} - \vec{v}(\vec{r}) - \vec{v}_i^{\prime}  \ ,
\label{eq:bond_rupture-3}
\end{equation}
where $\vec{v}_i^{\prime}$ and $\vec{v}_j^{\prime}$ are atomic velocities and $\vec{v}$ is the velocity of the medium. Let us denote
$\Delta \vec{v}_{ij}^{\prime} = \vec{v}_j^{\prime} - \vec{v}_i^{\prime}$. Then Eq.~(\ref{eq:bond_rupture-3}) can be written as:
\begin{equation}
\Delta \vec{v} =
\Delta \vec{v}_{ij}^{\prime} +
\left[ \vec{v}(\vec{r}+\vec{l}) - \vec{v}(\vec{r}) \right] \ .
\label{eq:bond_rupture-3-2}
\end{equation}
Assuming that the molecular bond is oriented parallel to the direction of the shock wave propagation, one derives
\begin{equation}
\vec{v}(\vec{r}+\vec{l}) - \vec{v}(\vec{r}) =
\left( \frac{\partial v}{\partial t} + v \, \frac{\partial v}{\partial r} \right) \, \frac{l}{v} \ .
\label{eq:bond_rupture-3-3}
\end{equation}
Using the Euler equation
\begin{equation}
\frac{\partial v}{\partial t} + v \, \frac{\partial v}{\partial r} = - \frac{1}{\rho} \, \frac{\partial P}{\partial r}
\label{eq:bond_rupture-3-4}
\end{equation}
where $\rho$ is the density of the medium, Eq.~(\ref{eq:bond_rupture-3-2}) is rewritten as
\begin{equation}
\Delta \vec{v} =
- \frac{1}{\rho} \, \frac{\partial P}{\partial r} \, \frac{l}{v} \, \vec{n}_r + \Delta \vec{v}_{ij}^{\prime} \ ,
\label{eq:bond_rupture-3-5}
\end{equation}
where $\vec{n}_r = \frac{\vec{r}}{r}$ is the unit vector.

Let us assume that the variation of atomic velocities $\Delta \vec{v}_{ij}^{\prime} \sim \sqrt{\frac{k_{B} T}{\mu}}$ caused by the thermal motion of atoms in the molecule is much smaller than the variation of the velocity of the medium caused by the shock wave propagation, i.e.
\begin{equation}
| \Delta \vec{v}_{ij}^{\prime} | \ll \frac{1}{\rho} \, \frac{\partial P}{\partial r} \, \frac{l}{v}  \ .
\label{eq:bond_rupture-3-6}
\end{equation}
In this case the condition for the bond rupture, Eq.~(\ref{eq:bond_rupture-6}), can be written in the form:
\begin{equation}
\frac{\mu}{2} \frac{l^2}{\rho^2 v^2}  \left( \frac{\partial P}{\partial r} \right)^2 \ge \Delta E \ .
\label{eq:bond_rupture-6-2}
\end{equation}

The energy barrier $\Delta E$ can be easily evaluated. Thus, equating the derivative of $U(r)$, Eq.~(\ref{eq:bond_rupture-1}), over $r$ to zero one derives
\begin{equation}
D_e \left[ -2\kappa \, e^{-2\kappa (r - r_0)} + 2 \kappa \, e^{-\kappa (r - r_0)} \right] - F = 0 \ .
\label{eq:bond_rupture-7}
\end{equation}
By solving this equation one obtains the values $\Delta r_1 = r_1 - r_0$ and $\Delta r_2 = r_2 - r_0$:
\begin{equation}
\Delta r_{1,2} = -\frac{1}{\kappa} \, \ln{ \left[ \frac{1}{2} \pm \frac{1}{2} \, \sqrt{1 - 2 \alpha}  \right] }
\label{eq:bond_rupture-8}
\end{equation}
where
\begin{equation}
\alpha = \frac{F}{\kappa \, D_e}
\label{eq:bond_rupture-9}
\end{equation}
is the dimensionless parameter.

Substituting Eq.~(\ref{eq:bond_rupture-8}) into Eq.~(\ref{eq:bond_rupture-2}) and performing simple algebraic transformations, one derives the expression for the energy barrier $\Delta E$ in the following form:
\begin{equation}
\Delta E = D_e \left[  \sqrt{1 - 2\alpha} +  \alpha \, \ln{ \frac{ 2 \alpha }{ \left( 1 + \sqrt{1 - 2 \alpha} \right)^2 } } \right] \ .
\label{eq:bond_rupture-10}
\end{equation}

Combining Eqs.~(\ref{eq. shock wave propagation}) and (\ref{eq:MSA_SW_front_pressure}), one obtains the relationship between the pressure at the shock wave front and the velocity of the medium caused by the shock wave propagation \cite{surdutovich2010shock}:
\begin{equation}
P(R) = \frac{2}{\gamma + 1} \rho \, v^2(R) \ .
\label{eq:bond_rupture-101}
\end{equation}
Substituting Eqs.~(\ref{eq:stretching_force}), (\ref{eq:bond_rupture-9}), (\ref{eq:bond_rupture-10}) and (\ref{eq:bond_rupture-101}) into Eq.~(\ref{eq:bond_rupture-6-2}), one derives the following condition for bond rupture:
\begin{equation}
\eta \, \frac{l}{P(R)} \, \left( \frac{\partial P}{\partial r} \right) \, \alpha  \ge
\sqrt{1 - 2\alpha} + \alpha \, \ln{ \frac{ 2 \alpha }{ \left( 1 + \sqrt{1 - 2 \alpha} \right)^2 } } \ ,
\label{eq:bond_rupture-11}
\end{equation}
where
\begin{equation}
\eta = \frac{1}{\gamma+1} \, \frac{\mu \, \kappa }{\pi a_0^2 \, \rho } \,
\end{equation}
is the dimensionless parameter. The left-hand side of Eq.~(\ref{eq:bond_rupture-11}) is a function of distance from the ion track, $r$, and the position of the shock wave front, $R$, at a given time moment $t$.

The parametric inequality (\ref{eq:bond_rupture-11}) is fulfilled in the region $\alpha \ge \bar{\alpha}$, where the threshold value $\bar{\alpha}$ is determined by equating the left- and right-hand sides of Eq.~(\ref{eq:bond_rupture-11}).
The pressure gradient $\frac{\partial P}{\partial r}$ in the vicinity  of the shock wave front can be related to $\frac{\partial P}{\partial R}$, that is the derivative of the pressure on the shock wave front $P(R)$, Eq.~(\ref{eq:MSA_SW_front_pressure}), with respect to the shock wave front radius $R$.
As demonstrated earlier for the carbon ion at the Bragg peak \cite{surdutovich2010shock} and derived in the present study for the heavier ions at the Bragg peak, the following relation applies:
\begin{equation}
\left. \frac{\partial P}{\partial r} \right|_{r = R} = \nu  \left| \frac{\partial P}{\partial R} \right| \ ,
\label{eq:pressure_gradient_dPdR}
\end{equation}
where the proportionality factor $\nu = 5.95$ is independent on ion's LET and time.
Differentiating Eq.~(\ref{eq:MSA_SW_front_pressure}) over $R$ and combining Eqs.~(\ref{eq:stretching_force}) and (\ref{eq:bond_rupture-9}), the condition defining the solution of Eq.~(\ref{eq:bond_rupture-11}) can be expressed as:
\begin{equation}
\frac{\pi \nu \beta^4}{\gamma + 1} \, \frac{ a_0^2 \, l}{\kappa \, D_e} \, \frac{S_e}{R^3} \ge \bar{\alpha} \ .
\label{eq:bond_rupture-102}
\end{equation}
From this expression one derives the threshold distance from the ion track, $R_{{\rm SW}}$, below which the bonds in the DNA backbone can be broken by the shock wave imposed thermomechanical stress:
\begin{equation}
R_{{\rm SW}} =  b \, S_e^{1/3} \ ,
\label{eq:bond_rupture-13}
\end{equation}
where the pre-factor $b$ reads as
\begin{equation}
b =  \left( \frac{\pi \nu \beta^4}{\gamma + 1} \,  \frac{a_0^2 \, l}{\kappa D_e \, \bar{\alpha}}  \right)^{1/3} \ .
\label{eq:bond_rupture-14}
\end{equation}
The distance $R_{{\rm SW}}$ depends on the parameters of a specific covalent bond ($D_e$ and $\kappa$) and on the ion's LET.

\begin{figure*}[t!]
\centering
\includegraphics[width=0.48\textwidth]{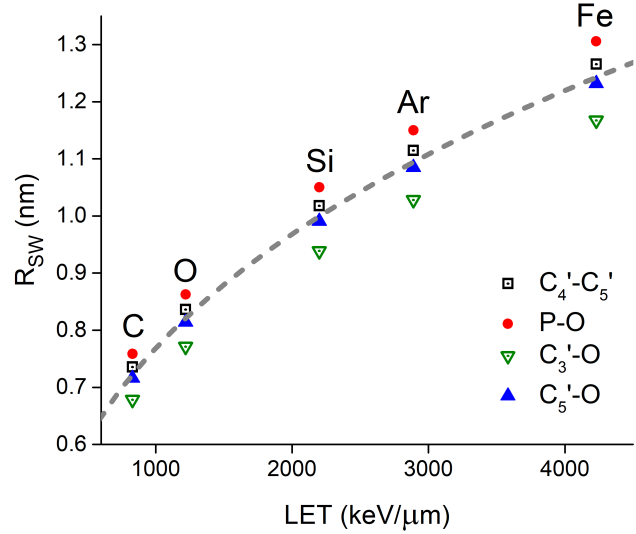}
\caption{The threshold distance $R_{{\rm SW}}$ for cleavage of the C$_3^{\prime}$--O, C$_4^{\prime}$--C$_5^{\prime}$, C$_5^{\prime}$--O and P--O bonds by the shock wave induced thermomechanical stress by the five studied ions at the Bragg peak region. The $R_{{\rm SW}}$ values are calculated according to Eq.~(\ref{eq:bond_rupture-13}) (symbols). The coefficient $b$, Eq.~(\ref{eq:bond_rupture-14}), varies in the range $(0.072 - 0.081)$~nm$^{4/3}$\,eV$^{-1/3}$ for the four bonds considered. Dashed line shows the $R_{{\rm SW}} = b \, S_e^{1/3}$ dependence with the average value $b = 0.077$~nm$^{4/3}$\,eV$^{-1/3}$. }
\label{fig:SW_forces}
\end{figure*}

\begin{table}[t!]
\centering
\caption{The critical distance from the ion track, $R_{{\rm SW}}$, below which covalent bonds in the DNA backbone can be broken by the shock wave induced thermomechanical stress in the vicinity of the Bragg peak. The $R_{{\rm SW}}$ values are given in nanometers.
The bottom line lists the corresponding threshold values of the external force $F$, Eq.~(\ref{eq:stretching_force}).}
\begin{tabular}{p{2.0cm}p{1.5cm}p{1.5cm}p{1.5cm}p{1.5cm}p{1.3cm}}
\hline
        &  C$_3^{\prime}$--O  &  C$_4^{\prime}$--C$_5^{\prime}$  &  C$_5^{\prime}$--O  &  P--O  &  average  \\
\hline
Carbon  &        0.68         &               0.73               &          0.71       &  0.76  &    0.72   \\
Oxygen  &        0.77         &               0.83               &          0.81       &  0.86  &    0.82   \\
Silicon &        0.94         &               1.01               &          0.99       &  1.05  &    1.00   \\
Argon   &        1.03         &               1.11               &          1.08       &  1.15  &    1.09   \\
Iron    &        1.16         &               1.26               &          1.23       &  1.30  &    1.24   \\
\hline
$F$~(nN) &        6.7         &                5.2               &           5.7       &   4.8  &     5.6   \\
\hline
\end{tabular}
\label{tab:Critical_distance_rupture}
\end{table}

Figure~\ref{fig:SW_forces} shows the dependence of $R_{{\rm SW}}$ on $S_e$ for the five studied ions in the Bragg peak region. Symbols show the $R_{{\rm SW}}$ values for cleavage of the C$_3^{\prime}$--O, C$_4^{\prime}$--C$_5^{\prime}$, C$_5^{\prime}$--O and P--O bonds in the DNA backbone, calculated according to Eq.~(\ref{eq:bond_rupture-13}) and (\ref{eq:bond_rupture-14}).
The coefficient $b$ varies in the range $(0.072 - 0.081)$~nm$^{4/3}$\,eV$^{-1/3}$ for the four bonds considered. The $R_{{\rm SW}} = b \, S_e^{1/3}$ dependence with the average value $b = 0.077$~nm$^{4/3}$\,eV$^{-1/3}$ is shown in Fig.~\ref{fig:SW_forces} by the dashed line.
The calculated values for $R_{{\rm SW}}$ for the C$_3^{\prime}$--O, C$_4^{\prime}$--C$_5^{\prime}$, C$_5^{\prime}$--O and P--O bonds and their average value for different LET values are listed in Table~\ref{tab:Critical_distance_rupture}.
The table lists also the corresponding threshold values of the external force $F$ at which the bond rupture becomes possible. The threshold values of $F$ are evaluated using Eq.~(\ref{eq:stretching_force}) with the pressure gradient $\frac{\partial P}{\partial r}$ being related to the derivative of the pressure at the shock wave front, $\frac{\partial P}{\partial R}$, via Eq.~(\ref{eq:pressure_gradient_dPdR}).

The results presented in Fig.~\ref{fig:SW_forces} and Table~\ref{tab:Critical_distance_rupture} indicate that the threshold distance from the ion's path for the bond rupture by the pressure gradient on the shock wave front varies from 0.7~nm for a carbon ion at the Bragg peak to 1.3~nm for an iron ion at the Bragg peak.
The estimated $R_{{\rm SW}}$ values are consistent with the results of MD simulations shown in Fig.~\ref{fig:probability of ssb}.
Note however that no strand breaks have been observed in the simulations for carbon and oxygen projectile ions for the default bond dissociation energies $D_e$ (Fig.~\ref{fig:probability of ssb}C), which can be attributed to a small number of simulated trajectories and low number of the events. Accounting for different possible orientations of the molecular bonds in the DNA backbone should also increase their average stability.
Simulations performed with the scaled bond dissociation energies $D_e/2$ and $D_e/6$ (Fig.~\ref{fig:probability of ssb}B and Fig.~\ref{fig:probability of ssb}A, respectively) indicate the formation of bond breaks in the DNA backbone by the carbon- and oxygen-ion induced shock wave within the range of distances from the ion track, which are consistent with the $R_{{\rm SW}}$ values determined by Eq.~(\ref{eq:bond_rupture-13}) and listed in Table~\ref{tab:Critical_distance_rupture}.

\subsection{Shock wave induced DNA lethal damage of cells irradiated with high-LET ions}
\label{sec:Results_cell-surv}

\begin{figure}[t!]
\centering
\includegraphics[width=0.85\textwidth]{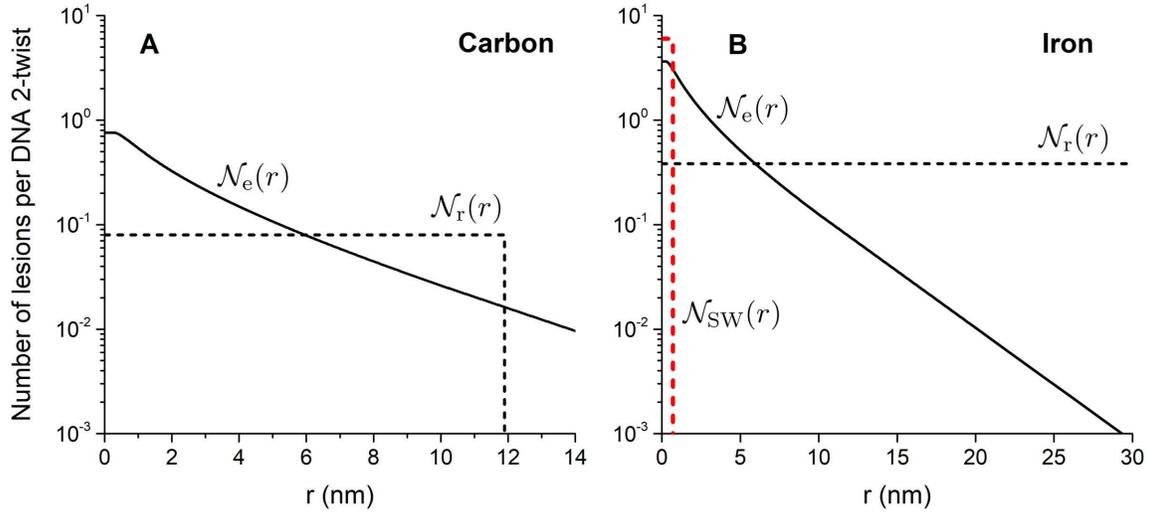}
\caption{Average number of simple lesions per DNA double twist due to a single carbon ion (\textbf{A}) and iron ion (\textbf{B}) at their Bragg peak energies, as a function of radial distance from the ion's path.
${\cal N}_{{\rm e}}(r)$ and ${\cal N}_{{\rm r}}(r)$ are the numbers of simple lesions produced by secondary electrons and free radicals, respectively.
${\cal N}_{{\rm SW}}(r)$ is the average number of lesions produced due to direct thermomechanical damage by the ion-induced shock wave.
The value ${\cal N}_{{\rm SW}} = 5.4$ at $r \leq R_{{\rm SW}} = 0.4$~nm was obtained from MD simulations, as summarized in Fig.~\ref{fig:probability of ssb} and Table~\ref{tab:Numbreaks_LET_De}. See the text for further details.}
\label{fig:num_lesions}
\end{figure}

Figure~\ref{fig:num_lesions} shows the average number of simple lesions per a DNA double twist as a function of radial distance from the ion's path for irradiation with a carbon ion (Fig.~\ref{fig:num_lesions}A) and with an iron ion (Fig.~\ref{fig:num_lesions}B) in the vicinity of the corresponding Bragg peaks.
The average number of simple lesions created by secondary electrons and free radicals
(${\cal N}_{{\rm e}}(r, S_e)$ and ${\cal N}_{{\rm r}}(r, S_e)$), is calculated according to Eqs.~(\ref{eq:MSA_numlesions_Ne}) and (\ref{eq:MSA_numlesions_Nr}), respectively.
The average number of lesions created by the shock wave induced thermomechanical stress of the DNA, ${\cal N}_{{\rm SW}}$, is taken from the MD simulations described in Sect.~\ref{sec:Number_bond_breaks_MD} (see Table~\ref{tab:Numbreaks_LET_De}). The number of breaks corresponds to the bond dissociation energies $D_e$ obtained from the DFT calculations \cite{friis_2020}.

As follows from the MD simulations (see Fig.~\ref{fig:probability of ssb} and Table~\ref{tab:Numbreaks_LET_De}) the thermomechanical stress by the carbon ion induced shock wave does not produce any lesions within the DNA double twist for the bond dissociation energies $D_e$. In the case of irradiation with a carbon ion, the lesions are created by secondary electrons, free radicals and other reactive species which are spread over the large distance range by the shock wave, see Fig.~\ref{fig:num_lesions}A.
This is in agreement with the results of earlier studies \cite{surdutovich2013biodamage, Surdutovich_2014_EPJD.68.353} which demonstrated that at the values of LET typical for a single carbon ion at the Bragg peak ($S_e = 830$~keV/$\mu$m), most of ion-induced DNA damage occurs via the chemical effects involving interactions of DNA molecules with secondary electrons, free radicals, solvated electrons, etc.
In contrast, the number of lesions produced by the thermomechanical stress caused by the iron ion induced shock wave outweighs the number of lesions produced by the chemical effects at distances $r \leq R_{{\rm SW}}$ from the ion's path, as shown in  Fig.~\ref{fig:num_lesions}B.
The analysis described below has been performed using the effective radius $R_{{\rm SW}} = 0.4$~nm for the iron ion at the Bragg peak, which is lower than the values reported in Sect.~\ref{sec:rupture_force}. One should stress that the model presented in Sect.~\ref{sec:rupture_force} gives the maximal values of $R_{{\rm SW}}$ for the five studied ions at the Bragg peak, corresponding to the ideal orientation of the molecular bond parallel to the direction of the shock wave propagation. Accounting for different orientations of the bonds in the DNA backbone with respect to the direction of a shock wave propagation should lead to lowering the $R_{{\rm SW}}$ values. The value $R_{{\rm SW}} = 0.4$~nm has been obtained by averaging the number of lesions produced due to direct thermomechanical damage by the iron ion-induced shock wave, ${\cal N}_{{\rm SW}}$, over the range of distances from the ion track to the principal axis of the DNA molecule, see Fig.~\ref{fig:probability of ssb}C.

On the basis of the non-reactive MD simulations and subsequent estimates for the energy deposited into the DNA backbone bonds, it was concluded earlier \cite{surdutovich2013biodamage} that the bond breaking due to the shock wave induced thermomechanical stress becomes dominant for ions heavier than argon, propagating in liquid water. This result is confirmed in the present study by means of reactive MD simulations.

\begin{figure}[t!]
\centering
\includegraphics[width=0.85\textwidth]{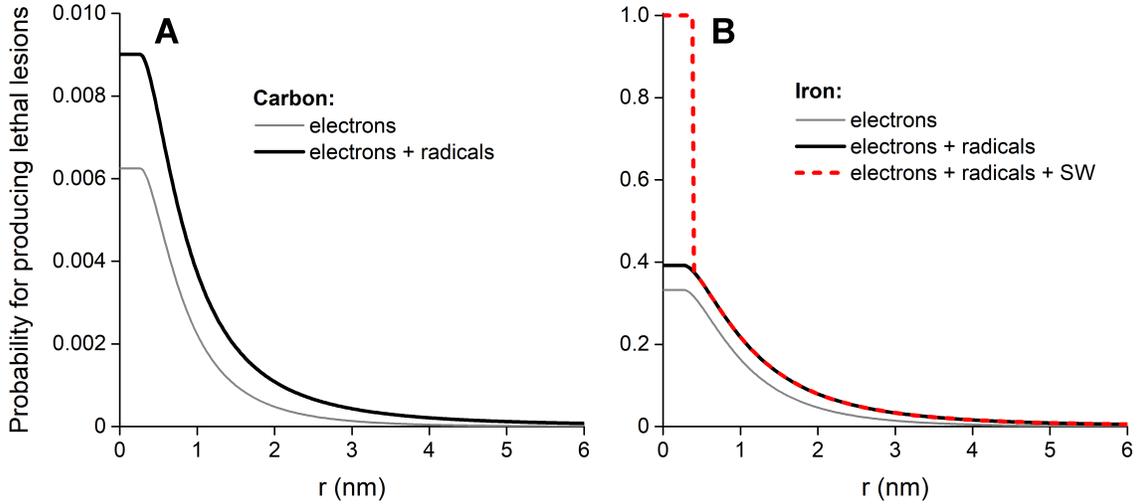}
\caption{Probability for producing lethal lesions in a DNA double twist as a function of radial distance from the ion's path for irradiation with a carbon ion (panel~A) and with an iron ion (panel~B) in the vicinity of the corresponding Bragg peaks.
Solid gray, solid black and dashed red curves show, respectively, the contribution of only secondary electrons, secondary electrons
and free radicals, as well as these agents together with the shock wave (SW) induced thermomechanical stress of the DNA.}
\label{fig:num_lesions-2}
\end{figure}

\begin{figure}[t!]
\centering
\includegraphics[width=0.48\textwidth]{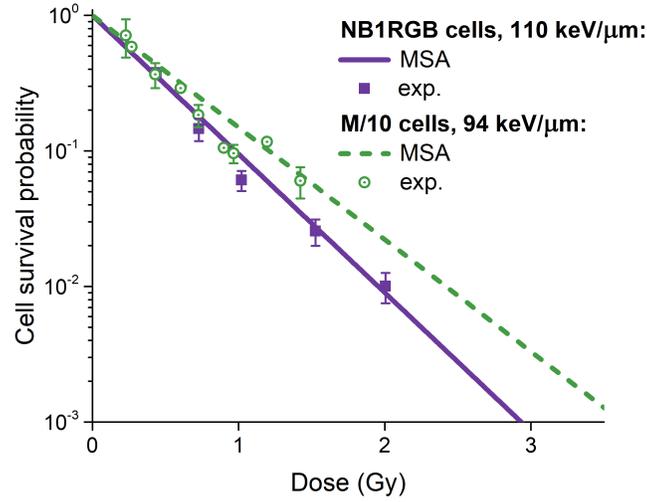}
\caption{Survival probability as a function of deposited dose for the normal tissue human fibroblast cell lines, NB1RGB and M/10, irradiated with carbon ions. Survival probabilities calculated within the MSA using Eqs.~(\ref{eq:MSA_prob_lethal_final})--(\ref{eq_cell_surv_prob}) at the indicated values of LET are shown with lines.
Experimental data for the NB1RGB \cite{Suzuki_2000_IJROBP.48.241} and M/10 \cite{Belli_2008_JRadiatRes.49.597} cells measured at a specific dose are shown by symbols.}
\label{fig:cell_surv_carbon}
\end{figure}

The calculated probabilities ${\cal P}_l(r, S_e)$ of producing the lethal DNA damage in a DNA double twist located at distance $r$ from the ion's path, Eqs.~(\ref{eq:MSA_prob_lethal_final}) and (\ref{eq:MSA_prob_lethal_final+SW}), are shown in Fig.~\ref{fig:num_lesions-2} for carbon and iron ions.
In the case of irradiation with iron ions (see Fig.~\ref{fig:num_lesions-2}B), accounting for the shock wave induced thermomechanical stress results in a significant increase of the probability of lethal DNA damage within the characteristic distance $r \le R_{{\rm SW}}$ from the ion track.
A conservative estimate for the number of bond breaks produced by the iron ion induced shock wave thermomechanical stress, corresponding to the largest bond dissociation energy $D_e$ (see Fig.~\ref{fig:probability of ssb} and Table~\ref{tab:Numbreaks_LET_De}), reveals that five or more bond breaks within the DNA double twist are created when the iron ion propagates at distances smaller than $R_{\rm SW} = 0.4$~nm from the principal axis of inertia of the DNA molecule.
The indicated number of breaks exceeds the minimal number of lesions needed to produce the lethal DNA damage, and hence the probability ${\cal P}_l(r, S_e) = 1$ at $r \le 0.4$~nm from the iron ion track. This means that even a single hit of a cell nucleus by a high-LET ion will be sufficient to inactivate the cell.

Figure~\ref{fig:cell_surv_carbon} shows survival probabilities for two human fibroblast cell lines irradiated
with carbon ions at high values of LET; the probabilities were evaluated within the MSA using Eqs.~(\ref{eq:MSA_prob_lethal_final})--(\ref{eq_cell_surv_prob}).
Lines show the survival curves obtained with accounting for the DNA damage produced by the secondary electrons, free radicals and the shock wave mechanism.
For carbon ion irradiation, the shock wave mechanism enhances transport of radicals and thus reduces their fast recombination thereby increasing the damaging effect of projectile ions. However, the direct thermomechanical DNA damage by the shock wave plays a minor role in the case of carbon ion irradiation.
One should stress a good agreement of the calculated survival probabilities with experimental data
\cite{Suzuki_2000_IJROBP.48.241, Belli_2008_JRadiatRes.49.597}.
These calculations were performed using the range of shock wave driven propagation of reactive species, $R_{{\rm r}} = 11.9$~nm, which was determined from the reactive MD simulations described in Sect.~\ref{sec:SW_propagation} (see Table~\ref{tab:Critical_Pressure}).

\begin{figure*}[t!]
\centering
\includegraphics[width=0.85\textwidth]{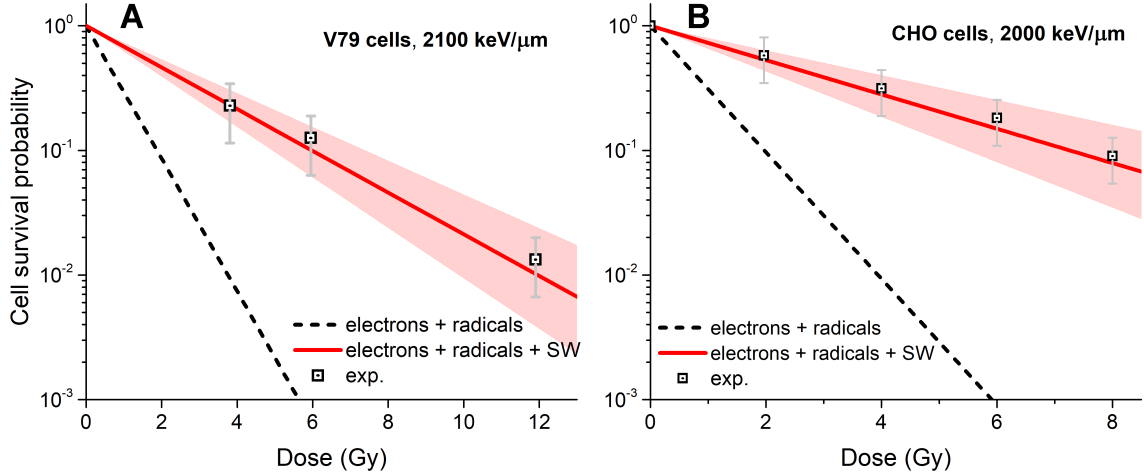}
\caption{Survival probability as a function of deposited dose for normal rodent cells, V79 and CHO, irradiated with iron ions at the indicated values of LET in the vicinity of the Bragg peak. Solid red lines show the probabilities calculated within the MSA framework with accounting for the shock-wave induced thermomechanical damage. Shaded areas illustrate variation of cell survival probabilities due to variation in the cell nucleus area (see text for details). Dashed lines show the cell survival probabilities calculated with accounting for the DNA damage produced only by secondary electrons and free radicals. Symbols denote experimental data for irradiation of the V79 \cite{Hirayama_2009_RadRes.171.212} and CHO \cite{Mehnati_2005_JRadiatRes.46.343} cells.}
\label{fig:cell_surv_iron}
\end{figure*}

The shock wave mechanism plays even bigger role in producing lethal damage to cells by high-LET ions as demonstrates in Fig.~\ref{fig:num_lesions-2}B.
Figure~\ref{fig:cell_surv_iron} shows survival probabilities for two normal rodent cells, V79 and CHO, irradiated with high-LET iron ions in the vicinity of the Bragg peak.
Solid red lines show the probabilities calculated with accounting for the shock-wave induced thermomechanical damage.
These probabilities were calculated within the MSA using the number of lethal lesions $Y_l$, defined by Eq.~(\ref{eq:yield_lethal_nucleus+SW}), and $R_{{\rm SW}} = 0.4$~nm, as discussed above.
Dashed black lines show the probabilities calculated with accounting for DNA damage produced only by secondary electrons and free radicals.
It is apparent that for the irradiation with high-LET ions the shock wave induced thermomechanical stress of the DNA has a significant impact on the cell survival probabilities.
If this mechanism is not taken into consideration, the calculated survival probabilities deviate by orders of magnitude from the experimental values \cite{Hirayama_2009_RadRes.171.212, Mehnati_2005_JRadiatRes.46.343}.
Indeed, according to Eqs.~(\ref{eq:MSA_eq03})--(\ref{eq_cell_surv_prob}), the number of lethal lesions $Y_l$ produced by secondary electrons and free radicals in a cell nucleus grows with an increase of LET.
As a consequence, the slope of cell survival curves would monotonically increase with an increase of LET.
This behavior contradicts with experimentally observed phenomenon known as the ``overkill'' effect, which manifests itself when cells are irradiated with high-LET ions.
At higher LET a given dose can be delivered with the smaller number of ions. This increases chances that some cells remain non-targeted, i.e. the cell survival probability should increase.
This leads to a less steep dependence of cell survival probability on the deposited dose \cite{Linz2012_IonBeams}.

Different approaches have been adopted in existing radiobiological models to account for the ``overkill'' effect. For instance, empirical saturation corrections due to non-Poisson distribution of lethal lesions in the cell nucleus were introduced in the commonly used LEM and MKM models to describe the radiobiological response to high-LET irradiation \cite{Hawkins_2003_RadiatRes.160.61, Kase_2008_PMB.53.37}.
In contrast to other models, the MSA describes quantitatively the ``overkill'' effect through accounting for the shock wave induced thermomechanical stress of the DNA.

As follows from Eq.~(\ref{eq:yield_lethal_nucleus+SW}), the quantification of the number of lethal lesions
produced by the ion-induced shock wave in a cell requires data on nucleus area for a particular cell line.
Solid red curves in Figs.~\ref{fig:cell_surv_iron}(A,B) are obtained with the values $A_n(\textrm{V79}) = 88$~$\mu$m$^2$ and
$A_n(\textrm{CHO}) = 127$~$\mu$m$^2$ taken, respectively, from the experimental studies \cite{Weyrather_1999_IJRB.75.1357,Konishi_2005_JRadiatRes.46.415}.
As it was reported by Konishi \textit{et al.} \cite{Konishi_2005_JRadiatRes.46.415}, the distribution of nucleus areas for the CHO cells
is characterized by a rather broad Gaussian-like profile, and the measured nucleus areas varies from about 80~$\mu$m$^2$ up to 160~$\mu$m$^2$ with the average value of 127~$\mu$m$^2$.
The variation of the calculated cell survival probabilities related to the variation of the nucleus size is illustrated in Fig.~\ref{fig:cell_surv_iron} by the shaded areas.
Note also that no data on the experimental uncertainties of the measured cell survival probabilities were provided in the earlier experimental studies \cite{Weyrather_1999_IJRB.75.1357, Konishi_2005_JRadiatRes.46.415}. Therefore, the characteristic uncertainties for the cells irradiated at doses up to about 10~Gy have been estimated based on the typical experimental uncertainties arising in such measurements with carbon ions (Fig.~\ref{fig:cell_surv_carbon}). The estimated uncertainties for the iron ion irradiation are shown in Fig.~\ref{fig:cell_surv_iron} by gray color.
One may thus conclude that, within the experimental uncertainties, the calculated survival probabilities for cells irradiated with iron ions are in a very good agreement with the experimental results \cite{Hirayama_2009_RadRes.171.212, Mehnati_2005_JRadiatRes.46.343}.
This agreement provides a strong experimental evidence for the biodamage effects caused by ion induced shock waves upon irradiation of biological targets with high-LET ions.

\section{Conclusions}

The thermomechanical stress of the DNA molecule caused by the ion-induced shock wave was explored using the reactive molecular dynamics (MD) simulations performed by means of high-performance computing. Five projectile ions with different values of LET, ranging from carbon to iron at the Bragg peak energies in liquid water, were considered. The number of bond breaks in the DNA backbone was systematically evaluated for each projectile ion as a function of bond dissociation energy and the distance from the ion's path to the principal axis of inertia of the DNA molecule.

Reactive MD simulations revealed that argon and, especially, iron ions induce rupture of multiple bonds in a DNA double twist containing 20 DNA base pairs. The DNA damage produced in segments of such size lead to complex irreparable lesions in a cell \cite{Surdutovich_2014_EPJD.68.353, verkhovtsev2016multiscale, schipler2013dna}. This makes the thermomechanical stress of the DNA molecule caused by the ion-induced shock wave the dominant mechanism of complex DNA damage at the high-LET ion irradiation.
In contrast, the shock wave induced by lighter ions, such as carbon and oxygen, causes only a few isolated bond breaks within a DNA double twist, but plays an important role in the transport of reactive species to larger distances away from the ion track.

A detailed theory for evaluating the DNA damage caused by ions at high-LET was formulated and integrated into the multiscale approach to the physics of radiation damage with ions (MSA). The theoretical analysis revealed that a single high-LET ion hitting a cell nucleus is sufficient to produce highly complex, lethal damages to a cell by the shock wave induced thermomechanical stress.
Using the parameters of the ion-induced shock wave propagation in liquid water, obtained numerically from MD simulations,
survival probabilities of cells irradiated with high-LET iron ions were evaluated by means of the MSA.
Accounting for the shock wave induced thermomechanical mechanism of DNA damage within the MSA provides an explanation for the ``overkill'' effect observed experimentally in the dependence of cell survival probabilities on the radiation dose delivered with iron ions.
A good agreement of the calculated cell survival probabilities with experimental data obtained for the cell irradiation with iron ions provides strong experimental evidence of the ion-induced shock wave effect and the related mechanism of radiation damage in cells.

\section*{Acknowledgements}

The authors are grateful for financial support from the Lundbeck Foundation, Danish Councils for Independent Research, the Volkswagen Foundation (Lichtenberg Professorship to IAS), the German Research Foundation DFG (Projects no. GRK1885, SFB 1382 and 415716638), and the European Union's Horizon 2020 research and innovation programme -- the Radio-NP project (GA 794733) within the H2020-MSCA-IF-2017 call (Individual Fellowship to AVV).
Computational resources for the simulations were provided by the DeiC National HPC Center, SDU and the CARL Cluster at
the Carl-von-Ossietzky University Oldenburg, which is supported by the DFG and the Ministry for Science and Culture of Lower Saxony.


\bibliography{bibliography}

\end{document}